\documentclass[letterpaper,12pt]{article}

\usepackage{geometry}    
\usepackage{amsmath, amssymb, amsthm, amsfonts}  
\usepackage{graphicx, subcaption}  
\usepackage{natbib}  
\usepackage{enumerate, paralist}  
\usepackage{arydshln, umoline}  
\usepackage{appendix}  
\usepackage{newtxtext, newtxmath}  
\usepackage{setspace}  
\usepackage{mathtools}  
\usepackage{physics}  
\usepackage{color}  
\usepackage{comment}  
\usepackage{tikz}

\newcommand{\calA}{\mathcal{A}}

\newcommand{\calE}{\mathcal{E}}

\newtheorem{theorem}{Theorem}
\newtheorem{lemma}{Lemma}
\newtheorem{proposition}{Proposition}
\newtheorem{corollary}[theorem]{Corollary}
\newtheorem{claim}{Claim}

\theoremstyle{definition}
\newtheorem{definition}{Definition}

\theoremstyle{remark}
\newtheorem{remark}{Remark}

\renewcommand{\baselinestretch}{1.5}  

\begin{document}
\title{Incomplete Information Robustness\thanks{This work is supported by Grant-in-Aid for Scientific Research Grant
Number 18H05217 and National Science Foundation Grant \#2001208. We
are grateful for the comments of Rafael Veiel, Satoru Takahashi, Takashi Kunimoto, and
seminar participants at the 12th Econometric Society World Congress, GAMES 2020, Singapore Management University, the University of Tokyo, and Keio University. }}
\author{Stephen Morris\thanks{Department of Economics, Massachusetts Institute of Technology.}\and
Takashi Ui\thanks{Department of Economics, Kanagawa University and Hitotsubashi Institute for Advanced Study.}}
\date{September 2026}

\maketitle
\global\long\def\baselinestretch{1.5}

\begin{abstract}
Consider an analyst who models a strategic situation using an incomplete information game. The true game may involve correlated, duplicated belief hierarchies, but the analyst lacks knowledge of the correlation structure and can only approximate each belief hierarchy.  To make predictions in this setting, the analyst uses belief-invariant Bayes correlated equilibria (BIBCE) and seeks to determine which of them are justifiable. We address this question by introducing the notion of robustness: a BIBCE is robust if, for every nearby incomplete information game, there exists a BIBCE close to it. Our main result provides a sufficient condition for robustness using a generalized potential function.  In a supermodular potential game, a robust BIBCE is a Bayes Nash equilibrium, whereas this need not hold in other classes of games.
\end{abstract}

\thispagestyle{empty}  

\newpage
\setcounter{page}{1}
\global\long\def\baselinestretch{1.5}%

\section{Introduction}
\label{Introduction}

When modeling a strategic situation as a game of incomplete information, an analyst must specify the types of players. Types can be decomposed into Mertens-Zamir belief hierarchies and an individually uninformative correlating device \citep{liu2015}, as different types may have correlated, duplicated belief hierarchies \citep{elypeski2006,dekeletal2007}. This decomposition presents two key challenges. First, accurately identifying the true belief hierarchies can be difficult. Second, even if the belief hierarchies are correctly identified, the correlating device may remain unknown.

If these challenges cannot be resolved, what predictions about players' behavior are justifiable? We address this question by considering an analyst who has only approximate knowledge of belief hierarchies and lacks information about their potential duplication and correlation. The analyst assumes that, while the true model may differ from her model, it lies within its neighborhood.  If a particular outcome of the analyst's model is qualitatively different from any outcome of some nearby game, as illustrated below, that outcome may not be justifiably adopted as a prediction. Then only outcomes that are (approximately) consistent with some outcomes of all nearby games can be considered justifiable, and we refer to such outcomes as robust.
 
\subsection*{Motivating example}
  
Let the analyst's model be an incomplete information game with two players $1$ and $2$; two actions $\alpha$ and $\beta$ for each player; and two payoff-relevant states $\theta_{1}$ and $\theta_{2}$ occurring with equal probability of $1/2$.\footnote{This game is discussed by \citet{liu2015}.}  Each player's set of types is a singleton; that is, it is common knowledge that each player has a uniform prior over $\{\theta_{1},\theta_{2}\}$. The payoffs are summarized in Table \ref{fig1}, where players would be better coordinated by choosing the same actions in state $\theta_{1}$ and different actions in state $\theta_{2}$.  This game has an infinite number of Bayes Nash equilibria (BNE) because the expected payoff for each player is $1/2$ under every action profile. 
\begin{table}[htbp]
\[
\begin{tabular}{l|cc}
 \ensuremath{\theta_{1}}  &  \ensuremath{\alpha}  &  \ensuremath{\beta} \\
\hline  \ensuremath{\alpha}  &  \ensuremath{1,1}  &  \ensuremath{0,0} \\
 \ensuremath{\beta}  &  \ensuremath{0,0}  &  \ensuremath{1,1} 
\end{tabular}\qquad\begin{tabular}{l|cc}
 \ensuremath{\theta_{2}}  &  \ensuremath{\alpha}  &  \ensuremath{\beta} \\
\hline  \ensuremath{\alpha}  &  \ensuremath{0,0}  &  \ensuremath{1,1} \\
 \ensuremath{\beta}  &  \ensuremath{1,1}  &  \ensuremath{0,0} 
\end{tabular}
\qquad
\begin{tabular}{l|cc}
 \ensuremath{\theta_{0}}  &  \ensuremath{\alpha}  &  \ensuremath{\beta} \\
\hline  \ensuremath{\alpha}  &  \ensuremath{1,0}  &  \ensuremath{1,1} \\
 \ensuremath{\beta}  &  \ensuremath{0,1}  &  \ensuremath{0,0} 
\end{tabular}
\]
\caption{Payoff matrices under \ensuremath{\theta_{1}}, \ensuremath{\theta_{2}}, and \ensuremath{\theta_{0}}}
    \label{fig1}
\end{table}

We construct a nearby game with a unique BNE, which is qualitatively different from any BNE in the analyst's model. Let $\theta_0$ be an additional payoff-relevant state associated with the rightmost payoffs in Table \ref{fig1}, where player 1 has a dominant action $\alpha$, and player 2's best response is not to match player 1's action. 
Let $n\in \{0,1,2,\ldots\}$ be a random variable drawn from the geometric distribution $\Pr(n=k)=\varepsilon(1-\varepsilon)^k$, which determines a payoff-relevant state as follows: $\theta_0$ occurs if $n=0$, $\theta_1$ occurs if $n\geq 1$ is odd, and $\theta_2$ occurs if $n\geq 2$ is even.

The types of players are defined by information partitions with respect to $n$, analogous to the email game \citep{rubinstein1989}. Player $1$'s partition is  $\{\{0\},\{1,2\},\{3,4\},\{5,6\},\ldots\}$; player~$2$'s partition is $\{\{0,1\},\{2,3\},\{4,5\},\{6,7\},\ldots\}$.  In the limit as $\varepsilon \to 0$, each player believes that either $\theta_1$ or $\theta_2$ occurs with equal probability $1/2$. The limit belief coincides with the belief in the analyst's model, so this game qualifies as a nearby game.

For any $\varepsilon>0$, we can iteratively eliminate dominated actions as follows.
\begin{itemize}
\item For player $1$ with $\{0\}$, $\alpha$ is a dominant action because the state is $\theta_0$. 
\item For player $2$ with $\{0,1\}$, $\beta$ is the unique best response because the state is $\theta_0$ with probability $1/(2-\varepsilon)$, and player $1$ chooses $\alpha$ with probability at least $1/(2-\varepsilon)$. 
\item For player $1$ with $\{1,2\}$, $\beta$ is the unique best response because the state is $\theta_1$ with probability $1/(2-\varepsilon)$, and player $2$ chooses $\beta$ with probability at least $1/(2-\varepsilon)$. 
\end{itemize}

Repeating this, we obtain a unique BNE with the following action profiles that survive iterated elimination of dominated actions: $(\alpha,\beta)$ when $n\in\{4k\}_{k=0}^{\infty}$, $(\beta,\beta)$ when $n\in\{4k+1\}_{k=0}^{\infty}$, $(\beta,\alpha)$ when $n\in\{4k+2\}_{k=0}^{\infty}$, and $(\alpha,\alpha)$ when $n\in\{4k+3\}_{k=0}^{\infty}$.  In the limit as $\varepsilon\to 0$, players choose $(\alpha,\alpha)$ and $(\beta,\beta)$ with equal probability when $\theta_{1}$ occurs, and $(\alpha,\beta)$ and $(\beta,\alpha)$ with equal probability when $\theta_{2}$ occurs.  This behavior is summarized in Table~\ref{fig3}.

\begin{table}[htbp]
\[
\begin{tabular}{l|cc}
 \ensuremath{\theta_{1}}  &  \ensuremath{\alpha}  &  \ensuremath{\beta} \\
\hline  \ensuremath{\alpha}  &  \ensuremath{1/4}  &  \ensuremath{0} \\
 \ensuremath{\beta}  &  \ensuremath{0}  &  \ensuremath{1/4} 
\end{tabular}\qquad\begin{tabular}{l|cc}
 \ensuremath{\theta_{2}}  &  \ensuremath{\alpha}  &  \ensuremath{\beta} \\
\hline  \ensuremath{\alpha}  &  \ensuremath{0}  &  \ensuremath{1/4} \\
 \ensuremath{\beta}  &  \ensuremath{1/4}  &  \ensuremath{0} 
\end{tabular}
\]
\caption{Joint probabilities over actions and a state}
    \label{fig3}
\end{table}

The limit outcome in Table~\ref{fig3} is not a BNE of the analyst's model but a Bayes correlated equilibrium (BCE). Players receive recommendations to play one of their actions, randomly generated by the correlating device specified in Table~\ref{fig3}. Following these recommendations is optimal because the expected payoff is one, which is the maximum ex-post payoff. Furthermore, the recommended action is individually uninformative, as it does not alter the players' prior beliefs about the state.  Such a correlating device is said to be belief-invariant, and a BCE with a belief-invariant correlating device is referred to as a belief-invariant Bayes correlated equilibrium (BIBCE).    

\subsection*{Main contribution}

We develop a framework to study robust outcomes in incomplete information games and derive sufficient conditions for robustness.  The concept of robustness to small perturbations of information structures originated with \citet{kajiimorris1997a}, where the analyst's model is a complete information game.  We extend this concept to incomplete information games, addressing issues inherent to the incompleteness of information, such as potential duplication and correlation of belief hierarchies.

To formally define a nearby game, we first consider a game, called an elaboration, that has a collection of belief hierarchies that coincides with that of the analyst's model, but may contain correlated, duplicated belief hierarchies.	
A nearby game is defined as a game in which the belief hierarchies are approximately those in an elaboration, and is referred to as an $\varepsilon$-elaboration, where $\varepsilon$ measures the discrepancies. 

We adopt BIBCE as a candidate for the analyst's prediction because, in the motivating example, the BIBCE in Table \ref{fig3} is the only candidate for a robust outcome. 
A set of BIBCE is said to be robust 
if every $\varepsilon$-elaboration has a BIBCE that is close to some BIBCE in that set for sufficiently small $\varepsilon$. If the set consists of a single BIBCE, we simply say that the BIBCE is robust.

To derive a sufficient condition for robust BIBCE, we introduce a generalized potential function \citep{morrisui2005}  for an incomplete information game. 
This function is defined on the Cartesian product of the set of states and a covering of each player's action set, i.e., a collection of action subsets whose union equals the action set. It contains information about each player's preferences over these action subsets. A potential function \citep{monderershapley1996} is a special case in which each covering consists of singletons.
	
A generalized potential function is associated with belief-invariant correlating devices that assign each player an action subset as a recommendation to choose an action from that subset. An important belief-invariant correlating device is one designed to maximize the expected value of a generalized potential function. For any such correlating device, there exists a BIBCE in which each player chooses an action from the recommended subset. This BIBCE is referred to as a GP-maximizing BIBCE.

Our main result, Theorem \ref{main theorem}, establishes that the set of GP-maximizing BIBCE is robust. In particular, if a game is a potential game (i.e., a game that admits a potential function), the set of BIBCE that maximize the expected value of a potential function is robust. This implies that the BIBCE in the motivating example (shown in Table \ref{fig3}) is robust because the game is a potential game in which every player has the same payoff function. Recall that no BNE is robust in this example, even if it maximizes the expected value of a potential function over all BNE. This contrasts with robustness in complete information games, where a unique potential-maximizing Nash equilibrium is always robust \citep{ui2001,pram2019}. 

As the motivating example illustrates, BNE are vulnerable to belief-invariant correlations in nearby games. This naturally raises the question of under what conditions a BNE is robust.
Section \ref{Robust BIBCE of supermodular games} shows that if a game is a supermodular potential game with a unique potential-maximizing BNE, then this BNE is robust.  As a numerical example, we consider a binary-action supermodular game with two players, which is a global game \citep{carlssonvandamme1993a} with discretized signals and no dominance regions. This game has a robust BNE, where players always choose ex-post risk-dominant actions. Even in the absence of dominance regions, allowing a richer set of perturbations for the incomplete information game leads to the same outcome as the global game selection.



\subsection*{Related literature}

Our study builds on the literature addressing issues arising from the fact that type spaces \citep{harsanyi1967} can embed more correlations than those captured by belief hierarchies \citep{mertenszamir1985}. \citet{elypeski2006} provide an extended notion of belief hierarchies for two-player games and show that interim independent rationalizability depends on types solely through those hierarchies. \citet{dekeletal2007} introduce the concept of interim correlated rationalizability and demonstrate that types with identical belief hierarchies lead to the same set of interim-correlated-rationalizable outcomes.

While these papers treat correlations implicitly, \citet{liu2015} explicitly introduces correlating devices. He shows that every incomplete information game is equivalent to the conjunction of a non-redundant game, where different types have different hierarchies, and a belief-invariant correlating device. This result implies that a BNE of a game is equivalent to a BIBCE of the non-redundant game in terms of outcomes. 

The concept of BIBCE is introduced by \citet{forges2006} and \citet{lehreretal2010} under the additional restriction that action recommendations cannot depend on the state.  A BIBCE is a special case of a BCE, which is introduced by \citet{bergemannmorris2013,bergemannmorris2016} as an analyst's prediction when she cannot rule out players having additional information. The work of \citet{bergemannmorris2013,bergemannmorris2016} also supports the use of BIBCE in our paper as a candidate for an analyst's prediction under insufficient and imprecise information.\footnote{For further motivation to study BIBCE and its relationship to BCE, see \citet{liu2015}, \citet{bergemannmorris2017}, and \citet{bergemannmorrisveiel2024}. For its relation to other special cases of BCE, see \citet{bergemannmorris2019}.} 

The use of BIBCE is an important distinction between our robustness framework and that of \citet{kajiimorris1997a}, who use BNE as an equilibrium concept for nearby games. We allow correlated decisions in nearby games, which follows \citet{pram2019}. 
He shows that, when the equilibrium concept in nearby games is an agent-normal form correlated equilibrium, the robustness of \citet{kajiimorris1997a} and its weaker version of \citet{kajiimorris2020a} are equivalent.\footnote{The weaker version is used by \citet{ui2001}. The difference is shown by \citet{takahashi2020}.}   

Several sufficient conditions for the robustness of \citet{kajiimorris1997a} have been discussed.\footnote{See \citet{kajiimorris2020a,kajiimorris2020b} and \citet{oyamatakahashi2020} for recent developments.} 
\citet{morrisui2005} introduce a generalized potential function and show that an equilibrium maximizing the generalized potential function is robust. A potential function and a monotone potential function are special cases. 
In the case of binary-action supermodular games, \citet{oyamatakahashi2020} show that a monotone potential function provides not only a sufficient but also a necessary condition for the robustness. Our sufficient condition for robust BIBCE extends these insights. 

Building on the necessity argument in \citet{oyamatakahashi2020}, \citet{morrisetal2020} characterize ``smallest equilibrium'' implementation in binary-action supermodular games with incomplete information through information design. Their work extends arguments from the robustness literature  to incomplete information settings, an approach shared with our paper. However, their focus differs from ours. While \citet{morrisetal2020} consider arbitrary information structures to induce desirable actions in binary-action supermodular games, our paper examines small perturbations in information structures to assess the robustness of outcomes in general games.

\section{Model}
\label{Elaborations of incomplete information games}


\subsection{Incomplete information games}

Fix a finite set of players $I$ and a finite set of actions $A_{i}$ for each player $i\in I$.  An incomplete information game $(T,\Theta,\pi,u)$ consists of the following elements  (we simply refer to it as a game when there is no risk of confusion).
\begin{itemize}
\item $T=\prod_{i\in I}T_{i}$ is an at most countable set of type profiles,
where $T_{i}$ is the set of player $i$'s types. 
\item $\Theta=\prod_{i\in I}\Theta_{i}$ is an at most countable set of
payoff-relevant states, where player $i$'s payoff function is determined by the $i$-th component $\theta_i\in\Theta_i$.\footnote{The restriction to countable types and payoff-relevant states is necessary for the existence of Bayes Nash equilibria and the comparison with the prior literature. BIBCE exist on uncountable type spaces via an argument of \citet{stinchcombe2011} as discussed by \citet{bergemannmorrisveiel2024}, so our results that restrict attention to BIBCE would extend to uncountable type spaces.} 
\item $\pi\in\Delta(T\times\Theta)$ is a common prior. 
\item $u=(u_{i})_{i\in I}$ is a payoff function profile, where $u_{i}:A\times\Theta\to\mathbb{R}$
is player $i$'s payoff function such that $u_{i}(\cdot,\theta)=u_{i}(\cdot,\theta')$
if and only if $\theta_{i}=\theta_{i}'$. 
\end{itemize}

Hereafter, we use $C=\prod_{i\in I}C_{i}$, $C_{-i}=\prod_{j\neq i}C_{j}$, $C_{S}=\prod_{i\in S}C_{i}$, and $C_{-S}=\prod_{i\not\in S}C_{i}$ to denote the Cartesian products of $C_{1},C_{2},\ldots$ with generic elements $c\in C$, $c_{-i}\in C_{-i}$, $c_{S}\in C_{S}$, and $c_{-S}\in C_{-S}$, respectively.

Payoff functions are assumed to be bounded, i.e., $\sup_{i,a,\theta}|u_{i}(a,\theta)|<\infty$. For each $i\in I$, let $T_{i}^{*}\subseteq T_{i}$ and $\Theta_{i}^{*}\subseteq\Theta_{i}$ denote the sets of player $i$'s types and payoff-relevant states on the support of $\pi$, respectively: $T_{i}^{*}=\{t_{i}\in T_{i}\mid\pi(t_{i})>0\}$ and $\Theta_{i}^{*}=\{\theta_{i}\in\Theta_{i}\mid\pi(\theta_{i})>0\}$, where $\pi(t_{i})\equiv\sum_{t_{-i},\theta}\pi(t,\theta)$ and $\pi(\theta_{i})\equiv\sum_{t,\theta_{-i}}\pi(t,\theta)$ are the marginal probabilities. Player $i$'s belief is given by $\pi(t_{-i},\theta|t_{i})\equiv\pi(t,\theta)/\pi(t_{i})$ when his type is $t_{i}\in T_{i}^{*}$.

Let $\pi^{*}\in\Delta(T^{*}\times\Theta^{*})$ and $u_{i}^{*}:A\times\Theta^{*}\to\mathbb{R}$ denote the restriction of $\pi$ to $T^{*}\times\Theta^{*}$ and that of $u_{i}$ to $A\times\Theta^{*}$, respectively. Note that $(T^{*},\Theta^{*},\pi^{*},u^{*})$ is the minimum representation of $(T,\Theta,\pi,u)$ because every player with every type on the support of $\pi$ in $(T,\Theta,\pi,u)$ has the same belief and payoffs as those in $(T^{*},\Theta^{*},\pi^{*},u^{*})$. We say that two games $(T,\Theta,\pi,u)$ and $(\bar{T},\bar{\Theta},\bar{\pi},\bar{u})$ with the same set of players and the same set of actions are equivalent if they have the same minimum representation.

A decision rule is a mapping $\sigma:T\times\Theta\rightarrow\Delta(A)$, under which players choose an action profile $a\in A$ with probability $\sigma(a|t,\theta)$ when $(t,\theta)\in T\times\Theta$ is realized.  Let $\Sigma$ denote the set of all decision rules. A decision rule $\sigma$ together with a common prior $\pi$ determines a joint probability distribution $\sigma\circ\pi\in\Delta(A\times T\times\Theta)$ given by $\sigma\circ\pi(a,t,\theta)\equiv\sigma(a|t,\theta)\pi(t,\theta)$, which is referred to as a distributional decision rule. The set of all distributional decision rules, denoted by $\Sigma\circ\pi\equiv\{\sigma\circ\pi\in\Delta(A\times T\times\Theta)\mid\sigma\in\Sigma\}$, is readily shown to be a compact subset of a linear space $\{f\in\mathbb{R}^{A\times T\times\Theta}\mid\sum_{a,t,\theta}|f(a,t,\theta)|<\infty\}$
with the weak topology, which is metrizable.\footnote{See Lemma~\ref{prohorov} in the appendix for more details.} Each distributional decision rule corresponds to an equivalence class of decision rules, where $\sigma,\sigma'\in\Sigma$ are equivalent if $\sigma\circ\pi(a,t,\theta)=\sigma'\circ\pi(a,t,\theta)$ for all $(a,t,\theta)\in A\times T^{*}\times\Theta^{*}$. When we discuss the  topology of $\Sigma$, we identify $\Sigma$ with $\Sigma\circ\pi$ by regarding $\Sigma$ as the set of  equivalence classes and considering the isomorphism from the set of equivalence classes to $\Sigma\circ\pi$.

\subsection{Belief-invariant Bayes correlated equilibrium}

A decision rule $\sigma$ is said to be {\em obedient} for player
$i$ of type $t_{i}$ if 
\begin{equation}
\sum_{a_{-i},\,t_{-i},\,\theta}\sigma(a|t,\theta)\pi(t,\theta)u_{i}(a,\theta)\geq\sum_{a_{-i},\,t_{-i},\,\theta}\sigma(a|t,\theta)\pi(t,\theta)u_{i}((a_{i}',a_{-i}),\theta)  \label{obedient condition 2}
\end{equation}
for all $a_{i},a_{i}'\in A_{i}$; that is, this player cannot increase the
expected payoff by deviating from the action prescribed by the decision
rule. In particular, if a decision rule $\sigma$ is obedient for
every player of every type, then $\sigma$ is simply said to be obedient.
An obedient decision rule is referred to as a Bayes correlated equilibrium
(BCE) \citep{bergemannmorris2013,bergemannmorris2016}.

A decision rule is {\em belief-invariant} if $\sigma\left(\{a_{i}\}\times A_{-i}|\left(t_{i},t_{-i}\right),\theta\right)$
is independent of $(t_{-i},\theta)$ for each $a_{i}\in A_{i}$,
$t_{i}\in T_{i}$, and $i\in I$, or equivalently, there exists a player $i$'s strategy $\sigma_{i}:T_{i}\to\Delta(A_{i})$ such that $\sigma_{i}(a_{i}|t_{i})=\sigma\left(\{a_{i}\}\times A_{-i}|\left(t_{i},t_{-i}\right),\theta\right)$.
Belief-invariance implies that player $i$'s action does not reveal
any additional information to the player about the opponents' types
and the state. In other words, from the viewpoint of player $i$ who
observes $(a_{i},t_{i})$ as a random variable, $t_{i}$ is a sufficient statistic for
$(t_{-i},\theta)$. This property has played an important role in
the literature on incomplete information correlated equilibrium \citep{forges1993,forges2006,lehreretal2010,liu2015}.
Let $\Sigma^{BI}$ denote the set of all belief-invariant decision
rules. It is readily shown that $\Sigma^{BI}$ is a compact subset
of $\Sigma$.

Note that a strategy profile $(\sigma_{i})_{i\in I}$, where $\sigma_{i}:T_{i}\to\Delta(A_{i})$
is player $i$'s strategy, is a special case of a belief-invariant
decision rule given by $\sigma(a|t,\theta)=\prod_{i\in I}\sigma_{i}(a_{i}|t_{i})$.
A strategy profile $(\sigma_{i})_{i\in I}$ is said to be a Bayes
Nash equilibrium (BNE) if it is obedient. Clearly, a BNE is a special
case of a belief-invariant BCE (BIBCE). Because a BNE exists in our
setting with at most a countable number of states, types, and actions
\citep{milgromweber1985}, a BIBCE also exists.\footnote{A BIBCE is also a solution to a linear programming problem with a
countable number of variables and constraints. } It is easy to check that the set of all BIBCE has the following property.

\begin{lemma} The set of all BIBCE of $(T,\Theta,\pi,u)$ is a nonempty
convex compact subset of $\Sigma$ containing all BNE, where a convex
combination $\alpha\sigma+(1-\alpha)\sigma'$ for $\sigma,\sigma'\in\Sigma^{BI}$
and $\alpha\in(0,1)$ is given by 
$
(\alpha\sigma+(1-\alpha)\sigma')(a|t,\theta)=\alpha\sigma(a|t,\theta)+(1-\alpha)\sigma'(a|t,\theta)
$
for all $a\in A$ and $(t,\theta)\in T\times\Theta$. \end{lemma}

\subsection{Elaborations}

We introduce a correlating device that sends a signal to player $i$ from a set $M_{i}$, which is at most countable. 
The probability distribution of a signal profile
$m=(m_{i})_{i\in I}\in M\equiv\prod_{i\in I}M_{i}$ is given by a
mapping $\rho:T\times\Theta\to\Delta(M)$, under which $m\in M$ is
drawn with probability $\rho(m|t,\theta)$ when $(t,\theta)\in T\times\Theta$
is realized. This mapping $\rho$ is referred to as a communication
rule. Belief-invariance of a communication rule is defined similarly.
That is, $\rho$ is {belief-invariant} if $\rho\left(\{m_{i}\}\times M_{-i}|\left(t_{i},t_{-i}\right),\theta\right)$
is independent of $\left(t_{-i},\theta\right)$ for each $m_{i}\in M_{i}$,
$t_{i}\in T_{i}$, and $i\in I$, which implies that a signal $m_i$ does not reveal any additional information to player $i$ 
about the opponents' types and the state.

Combining a game $(T,\Theta,\pi,u)$ and a
communication rule $\rho$, we can construct another 
game $(\bar{T},\Theta,\bar{\pi},u)$ with the same sets of players
and actions such that $\bar{T}_{i}=T_{i}\times M_{i}$ for each $i\in I$
and $\bar{\pi}(\bar{t},\theta)=\pi(t,\theta)\rho(m|t,\theta)$ for
each $\bar{t}=((t_{i},m_{i}))_{i\in I}\in\bar{T}$ and $\theta\in\Theta$.
This game is referred to as the conjunction
of $(T,\Theta,\pi,u)$ and $\rho$. Note that player $i$ in $(\bar{T},\Theta,\bar{\pi},u)$
receives $m_{i}$ as well as $t_{i}$, where $t_{i}$ is drawn according
to $\pi$ and $m_{i}$ is drawn according to $\rho$, and if $\rho$
is belief-invariant, then this player's knowledge about $(t_{-i},\theta)$
is exactly the same as that in the original game $(T,\Theta,\pi,u)$.

We consider a game that is isomorphic
to the conjunction of $(T,\Theta,\pi,u)$ and a belief-invariant
communication rule, which is referred to as 
an elaboration of $(T,\Theta,\pi,u)$.

\begin{definition}\label{def: elaboration} A 
game $(\bar{T},\Theta,\bar{\pi},u)$ is an elaboration of $(T,\Theta,\pi,u)$
if there exist a belief-invariant communication rule $\rho$ and mappings
$\tau_{i}:\bar{T}_{i}\to T_{i}$ and $\mu_{i}:\bar{T}_{i}\to M_{i}$
for each $i\in I$ such that the mapping $\bar{t}_{i}\mapsto(\tau_{i}(\bar{t}_{i}),\mu_{i}(\bar{t}_{i}))$
restricted to $\bar{T}_i^{*}$ is one-to-one and 
\begin{equation}
\bar{\pi}(\bar{t},\theta)=\pi(\tau(\bar{t}),\theta)\rho(\mu(\bar{t})|\tau(\bar{t}),\theta)\text{ for all \ensuremath{\bar{t}\in\bar{T}}},\label{def:elabo}
\end{equation}
where $\tau(\bar{t})=(\tau_{i}(\bar{t}_{i}))_{i\in I}$ and $\mu(\bar{t})=(\mu_{i}(\bar{t}_{i}))_{i\in I}$.
\end{definition}

Note that if $(\bar{T},\Theta,\bar{\pi},u)$
is the conjunction of $(T,\Theta,\pi,u)$ and a belief-invariant communication
rule $\rho$, it is an elaboration of $(T,\Theta,\pi,u)$ with
$\tau_{i}$ and $\mu_{i}$ given by $\tau_{i}(t_{i},m_{i})=t_{i}$
and $\mu_{i}(t_{i},m_{i})=m_{i}$ for all $\bar{t}_{i}=(t_{i},m_{i})\in\bar{T}_{i}$
and $i\in I$.\footnote{Equation \eqref{def:elabo} implicitly requires that $\sum_{\bar{t}\in\tau^{-1}(t)}\rho(\mu(\bar{t})|t,\theta)=1$
for all $(t,\theta)\in T^{*}\times\Theta^{*}$ and $\tau(\bar{T}^{*})=T^{*}$
because $\bar{\pi}(\bar{T}\times\Theta)=1$. }

The following necessary and sufficient condition for an elaboration does not explicitly refer to a belief-invariant communication rule.

\begin{lemma}\label{lemma equivalence elaboration} A  game $(\bar{T},\Theta,\bar{\pi},u)$ is an elaboration
of $(T,\Theta,\pi,u)$ if and only if there exists
a mapping $\tau_{i}:\bar{T}_{i}\to T_{i}$ with $\tau_i(\bar{T}_{i})=T_i^*$ for each $i\in I$ such that 
\begin{gather}
\bar{\pi}(\tau^{-1}(t),\theta)=\pi(t,\theta)\text{ for all }(t,\theta)\in T^{*}\times\Theta^{*},\label{lemma 1:1}\\
\bar{\pi}(\tau_{-i}^{-1}(t_{-i}),\theta|\bar{t}_{i})={\pi}(t_{-i},\theta|\tau_{i}(\bar{t}_{i}))\text{ for all }\bar{t}_{i}\in\bar{T}_{i}^{*}
\text{ and }(t_{-i},\theta)\in T_{-i}^{*}\times\Theta^{*},\label{lemma 1:2}
\end{gather}
where $\tau_{i}^{-1}(t_{i})=\{\bar{t}_{i}\in\bar{T}_{i}\mid\tau_{i}(\bar{t}_{i})=t_{i}\}$,
$\tau^{-1}(t)=\prod_{i\in I}\tau_{i}^{-1}(t_{i})$, and $\tau_{-i}^{-1}(t_{-i})=\prod_{j\neq i}\tau_{j}^{-1}(t_{j})$.
The mapping $\tau=(\tau_i(\cdot))_{i\in I}$ is referred to as an elaboration mapping. 
\end{lemma}

Note that a type $\bar{t}_{i}\in\bar{T}_{i}$
in an elaboration has the same belief hierarchy over $\Theta$ as
that of a type $\tau_{i}(\bar{t}_{i})\in T_{i}$ in the original game
by \eqref{lemma 1:2}. Thus, two types $\bar{t}_{i},\bar{t}_{i}'\in\bar{T}_{i}$
with $\tau_{i}(\bar{t}_{i})=\tau_{i}(\bar{t}_{i}')$ also have the
same belief hierarchy over $\Theta$.

\subsection{Non-redundant games}

We define the notion of a non-redundant game,
where different types have different belief hierarchies. 
In $(T,\Theta,\pi,u)$, we say that $t_{i},t_{i}'\in T_{i}$ have the same belief hierarchy if 
there exists another 
game $(T^{0},\Theta,\pi^{0},u)$ such that $(T,\Theta,\pi,u)$ is an elaboration of $(T^{0},\Theta,\pi^{0},u)$ with an elaboration mapping $\tau^{0}$
satisfying $\tau_{i}^{0}(t_{i})=\tau_{i}^{0}(t_{i}')$.\footnote{To state it formally, let $Z_{i}^{1}=\Delta(\Theta)$, $Z_{-i}^{k-1}=\prod_{j\in I\setminus\{i\}}Z_{j}^{k-1}$,
and $Z_{i}^{k}=\Delta(\Theta\times Z_{-i}^{1}\times\cdots\times Z_{-i}^{k-1})$
for $k\geq2$. For each $t_{i}\in T_{i}$, define $h_{i}^{k}(t_{i})\in Z_{i}^{k}$
by $h_{i}^{1}(\theta|t_{i})=\pi(\{(t_{-i},\theta)\mid t_{-i}\in T_{-i}\}|t_{i})$
for each $\theta\in\Theta$, and $h_{i}^{k}(\theta,\hat{h}_{-i}^{1},\ldots,\hat{h}_{-i}^{k-1}|t_{i})=\pi(\{(t_{-i},\theta)\mid h_{j}^{l}(t_{j})=\hat{h}_{j}^{l}\text{ for all \ensuremath{j\neq i} and \ensuremath{l\in\{1,\ldots,k-1\}}}\}|t_{i})$
for each $\theta\in\Theta$ and $\hat{h}_{-i}^{l}\in Z_{-i}^{l}$.
The belief hierarchy of $t_{i}$ is $h_{i}(t_{i})\equiv(h_{i}^{k}(t_{i}))_{k=1}^{\infty}$.
As shown by \citet{liu2015}, $h_{i}(t_{i})=h_{i}(t_{i}')$ if and
only if $(T,\Theta,\pi,u)$ is an elaboration of another game $(T^{0},\Theta,\pi^{0},u)$ with an elaboration mapping $\tau^{0}$ satisfying $\tau_{i}^{0}(t_{i})=\tau_{i}^{0}(t_{i}')$.} 
If all different types in $(T^{0},\Theta,\pi^{0},u)$ have different belief hierarchies, we say that $(T^{0},\Theta,\pi^{0},u)$ is a non-redundant representation of the original game $(T,\Theta,\pi,u)$.

\begin{definition} A game $(T^{0},\Theta,\pi^{0},u)$
is a non-redundant representation of $(T,\Theta,\pi,u)$ if it satisfies the following two conditions: (i) $(T,\Theta,\pi,u)$ is an elaboration of $(T^{0},\Theta,\pi^{0},u)$ with an elaboration mapping $\tau^{0}$; (ii) for $t_{i},t_{i}'\in T_{i}^{*}$, $\tau_{i}^{0}(t_{i})=\tau_{i}^{0}(t_{i}')$
if and only if $t_{i}$ and $t_{i}'$ have the same belief hierarchy.
If $(T,\Theta,\pi,u)$ is a non-redundant representation of itself, then we simply
say that $(T,\Theta,\pi,u)$ is a non-redundant game. \end{definition}

A non-redundant game\footnote{We can characterize a non-redundant game using belief hierarchies. Let $[t_{i}]\equiv\{t_{i}'\in T_{i}\mid h_{i}(t_{i})=h_{i}(t_{i}')\}$
and $[t]=([t_{i}])_{i\in I}$. Consider $(T^{0},\Theta,\pi^{0},u)$
with $T^0=\{[t]\mid t\in T\}$ and $\pi^{0}([t],\theta)=\pi(\{(t',\theta)\,|\,t'\in[t]\})$.
Then, it is straightforward to show that $(T^{0},\Theta,\pi^{0},u)$
is a non-redundant game of $(T,\Theta,\pi,u)$.} will serve as the analyst's model in the next section.
\citet{liu2015} shows that if two games have the same set of belief hierarchies and one is non-redundant, then the other is an elaboration of the non-redundant game.\footnote{\citet{liu2015} assumes a finite type space.}

A BIBCE of a non-redundant game can be understood as a BNE of its elaboration. 
To see this, let $(\bar{T},\Theta,\bar{\pi},u)$ be an elaboration of $({T},\Theta,{\pi},u)$ with $\tau$.
For a decision rule $\sigma$ of $({T},\Theta,{\pi},u)$ and a decision
rule $\bar{\sigma}$ of $(\bar{T},\Theta,\bar{\pi},u)$, we say that
$\sigma$ and $\bar{\sigma}$ are outcome equivalent if both decision rules induce the same probability distribution over $A\times T\times \Theta$: 
\begin{equation}
\sigma\circ\pi(a,t,\theta)=
\bar{\sigma}\circ\bar\pi(a,\tau^{-1}(t),\theta)
\label{decision rule correspondence 2}
\end{equation}
for all $(a,t,\theta)\in A\times T\times\Theta$, where 
$\bar{\sigma}\circ\bar\pi(a,\tau^{-1}(t),\theta)=\sum_{\bar{t}\in\tau^{-1}(t)}\bar{\sigma}\circ\bar{\pi}(a,\bar{t},\theta)$. 
In particular, we say that $\sigma$ and $\bar{\sigma}$ are equivalent if 
\begin{equation}
\sigma(a|\tau(\bar{t}),\theta)=\bar{\sigma}(a|\bar{t},\theta)\notag 
\end{equation}
for all $(a,\bar{t},\theta)\in A\times\bar{T}\times\Theta$. Clearly,
if $\sigma$ and $\bar{\sigma}$ are equivalent, then they are outcome
equivalent, but not vice versa.

The following two lemmas are due to \citet{liu2015}. 
The first lemma shows that the set of BIBCE of $(T,\Theta,\pi,u)$ coincides with the set of
BNE of all elaborations of $(T,\Theta,\pi,u)$ in terms of outcome
equivalence. 
This implies that if the analyst uses a non-redundant game as her model but allows for all elaborations as the true model, then each BIBCE serves as a candidate for her prediction.

\begin{lemma}\label{elaboration BIBCE proposition}
A decision rule $\sigma$ is a BIBCE of $(T,\Theta,\pi,u)$ if and
only if there exists an elaboration $(\bar{T},\Theta,\bar{\pi},u)$
with an elaboration mapping $\tau$ and a BNE $\bar{\sigma}=(\bar{\sigma}_{i})_{i\in I}$
such that $\sigma$ and $\bar{\sigma}$ are outcome equivalent, i.e.,
for all $(a,t,\theta)\in A\times T\times\Theta$, 
\begin{equation}
\sigma(a|t,\theta)\pi(t,\theta)={\sum_{\bar{t}\in\tau^{-1}(t)}\prod_{i\in I}\bar{\sigma}_{i}(a_{i}|\bar{t}_{i})\bar{\pi}(\bar{t},\theta)}.\notag 
\end{equation}
\end{lemma}

Moreover, the set of BIBCE of $(T,\Theta,\pi,u)$ also coincides with the set
of BIBCE of an arbitrary elaboration of $(T,\Theta,\pi,u)$ in terms
of outcome equivalence, as shown by the second lemma.
Thus, once BIBCE is adopted as the equilibrium concept for analysis, a non-redundant game suffices for consideration, since every elaboration has the same set of BIBCE.

\begin{lemma}\label{elaboration BIBCE proposition 2} Let $(\bar{T},\Theta,\bar{\pi},u)$
be an elaboration of $({T},\Theta,{\pi},u)$. 
If $\sigma$ is a BIBCE of $({T},\Theta,{\pi},u)$ and a decision
rule $\bar{\sigma}$ of $(\bar{T},\Theta,\bar{\pi},u)$ is equivalent
to $\sigma$, then $\bar{\sigma}$ is a BIBCE of $(\bar{T},\Theta,\bar{\pi},u)$.
If $\bar{\sigma}$ is a BIBCE of $(\bar{T},\Theta,\bar{\pi},u)$ and
a decision rule $\sigma$ of $({T},\Theta,{\pi},u)$ is outcome equivalent
to $\bar{\sigma}$, then $\sigma$ is a BIBCE of $({T},\Theta,{\pi},u)$.
\end{lemma}

\section{Robustness}
\label{varepsilon-Elaborations and robustness}

This section introduces a framework to construct nearby games of the analyst's model using elaborations and provides a formal definition of robust BIBCE. 


Let $(T,\Theta,\pi,u)$ be a non-redundant game, which we regard as the analyst's model. The analyst believes that a belief-invariant communication rule may exist but has no information about it. Thus, the analyst predicts players' behavior using the outcomes of BNE of elaborations.  By Lemma \ref{elaboration BIBCE proposition},
such outcomes are BIBCE of $(T,\Theta,\pi,u)$. 

We further assume that the analyst believes that the true 
game is in a ``neighborhood'' of elaborations of the analyst's model. We call such
a nearby game an $\varepsilon$-elaboration,
which is approximately an elaboration of $(T,\Theta,\pi,u)$.
An elaboration is a special case of  an $\varepsilon$-elaboration with $\varepsilon=0$ (see Figure \ref{fig:elaborations}).

\begin{figure}[t]
\centering
\begin{tikzpicture}[scale=0.42]

    \draw[line width=1.2pt] (0,-1.1) ellipse (5.0cm and 2.8cm);

    \draw[line width=1.2pt] (0,-1.9) ellipse (4.0cm and 2.0cm);

    \draw[line width=1.2pt] (0,-2.6) ellipse (3.1cm and 1.3cm);

    \node at (0,0.9) {\small $\varepsilon$-elaborations};
    \node at (0,-0.64) {\small elaborations};
    \node[align=center] at (0,-2.5) {\small analyst's model
    };
\end{tikzpicture}
\caption{The analyst's model is nested within elaborations and $\varepsilon$-elaborations.}
\label{fig:elaborations}
\end{figure}

\begin{definition}\label{def: 2} For $\varepsilon\geq0$, a  game $(\bar{T},\bar{\Theta},\bar{\pi},\bar{u})$ is an
{$\varepsilon$-elaboration} of $(T,\Theta,\pi,u)$ if the following
requirements are satisfied. 
\begin{enumerate}
\item $\Theta^{*}\subseteq\bar{\Theta}$ and $\bar{u}_{i}(\cdot,\theta)=u_{i}(\cdot,\theta)$
for all $\theta\in\Theta^{*}$. 
\item $\bar{\pi}(\bar{T}^{\sharp})\geq1-\varepsilon$, where $\bar{T}^{\sharp}=\prod_{i\in I}\bar{T}_{i}^{\sharp}$
and $\bar{T}_{i}^{\sharp}=\{\bar{t}_{i}\in\bar{T}_{i}\mid\bar{\pi}(\Theta_{i}^{*}\times\bar{\Theta}_{-i}|\bar{t}_{i})=1\}$. 
\item There exist $\tau_{i}:\bar{T}_{i}\to T_{i}$ with $\tau_i(\bar{T}_{i})=T_i^*$ and $\bar{T}_{i}^{\flat}\subseteq\bar{T}_{i}$
with $\bar{\pi}(\bar{T}_{i}^{\flat})\geq1-\varepsilon$ satisfying
\begin{gather}
\sup_{E\subseteq T\times\Theta}\Big|\sum_{(t,\theta)\in E}\bar{\pi}(\tau^{-1}(t),\theta)-\sum_{(t,\theta)\in E}\pi(t,\theta)\Big|\leq\varepsilon,\label{def 3:1}\\
\sup_{E_{-i}\subseteq T_{-i}\times\Theta}\Big|\sum_{(t_{-i},\theta)\in E_{-i}}\bar{\pi}(\tau_{-i}^{-1}(t_{-i}),\theta|\bar{t}_{i})-\sum_{(t_{-i},\theta)\in E_{-i}}{\pi}(t_{-i},\theta|\tau_{i}(\bar{t}_{i}))\Big|\leq\varepsilon\label{def 3:2}
\end{gather}
 for all $\bar{t}_{i}\in\bar{T}_{i}^{\flat}$, where $\tau(\bar t)=(\tau_i(\bar t_i))_{i\in I}$  and  $\tau_{-i}(\bar t_{-i})=(\tau_j(\bar t_j))_{j\neq i}$. We call the mapping $\tau$ an elaboration mapping (with some abuse of terminology).\end{enumerate}
\end{definition}

By the first condition, the set of payoff-relevant states in an $\varepsilon$-elaboration
includes that in the original game. All players in an $\varepsilon$-elaboration have
the same payoff functions as those in the original 
game with probability greater than $1-\varepsilon$ by the second
condition (recall that $\theta_i\in\Theta_i$ determines player $i$'s payoff function). 
The third condition implies that an $\varepsilon$-elaboration is approximately an elaboration because, when $\varepsilon=0$, 
\eqref{def 3:1} and \eqref{def 3:2} reduce to 
\eqref{lemma 1:1} and \eqref{lemma 1:2} in Lemma \ref{lemma equivalence elaboration}.  
It is straightforward to verify that the nearby game in the motivating example in the introduction satisfies the conditions in Definition \ref{def: 2}.

\begin{remark}
\citet{kajiimorris1997a} define an $\varepsilon$-elaboration of a complete information game, which is $(T,\Theta,\pi,u)$ with $T$ and $\Theta$ being singletons. 
Their $\varepsilon$-elaboration is an incomplete information game that satisfies the first two conditions of Definition~\ref{def: 2}. 
Thus, in the case of complete information, any $\varepsilon$-elaboration under our definition is also an $\varepsilon$-elaboration in the sense of \citet{kajiimorris1997a}. 
See Section~\ref{KM comparison} for further details.
\end{remark}

We are now ready to define robustness.  A set of BIBCE of a non-redundant game is said to be robust if, for sufficiently small $\varepsilon > 0$, every $\varepsilon$-elaboration has a BIBCE that is close to some BIBCE in this set.\footnote{We can also consider robustness in ``redundant'' games. However, robust BIBCE in such games are outcome equivalent to those in non-redundant games, so it suffices to consider the non-redundant case.}

\begin{definition}\label{robust definition} 
Let $(T,\Theta,\pi,u)$
be a non-redundant game. A set of BIBCE of $(T,\Theta,\pi,u)$, $\mathcal{E}\subseteq\Sigma^{BI}$,
is robust if, for every $\delta>0$, there exists $\bar{\varepsilon}>0$
such that, for all $\varepsilon\leq\bar{\varepsilon}$, every $\varepsilon$-elaboration
$(\bar{T},\bar{\Theta},\bar{\pi},\bar{u})$ with an elaboration mapping $\tau$ has a BIBCE $\bar{\sigma}$ such that 
\begin{equation}
\sup_{E\subseteq A\times T^*\times\Theta}\left|\sum_{(a,t,\theta)\in E}\bar{\sigma}\circ{\bar{\pi}}(a,\tau^{-1}(t),\theta)-\sum_{(a,t,\theta)\in E}\sigma\circ\pi(a,t,\theta)\right|\leq\delta\label{condition1-0}
\end{equation}
for some $\sigma\in\mathcal{E}$, where $\bar{\sigma}\circ{\bar{\pi}}(a,\bar{t},\theta)=\bar{\sigma}(a|\bar{t},\theta)\bar{\pi}(\bar{t},\theta)$
and $\sigma\circ\pi(a,t,\theta)=\sigma(a|t,\theta)\pi(t,\theta)$. 
If $\calE=\{\sigma\}$ is a singleton, then $\sigma$ is said to be robust.
\end{definition}

To interpret \eqref{condition1-0}, recall 
Lemma \ref{elaboration BIBCE proposition 2}, which says
every $0$-elaboration has a BIBCE $\bar\sigma$ that is equivalent to 
a BIBCE $\sigma$ of $(T,\Theta,\pi,u)$, where \eqref{decision rule correspondence 2}
holds for all $(a,t,\theta)\in A\times T\times \Theta$.
Equation \eqref{condition1-0} means that \eqref{decision rule correspondence 2} holds approximately for sufficiently small $\varepsilon>0$ in the case of $\varepsilon$-elaborations.

\begin{remark}
Given the definition of robustness, the motivating example in Section \ref{Introduction} is summarized as follows and illustrates why we focus on BIBCE in our robustness concept. We have shown that there exists an $\varepsilon$-elaboration with a unique BNE. This BNE is close to the BIBCE in Table \ref{fig3} when $\varepsilon$ is small, but it is not close to any BNE of the analyst's model. This implies that the set of BNE for the analyst's model is not robust, so we adopt BIBCE for the analyst's model. We also adopt BIBCE for $\varepsilon$-elaborations because there exists an $\varepsilon$-elaboration whose BNE is not the BIBCE in Table \ref{fig3}. For instance, the analyst's model is a $0$-elaboration, where every BNE differs from the BIBCE in Table \ref{fig3}.
\end{remark}

\begin{remark}
Robustness is defined in terms of a distributional decision rule $\sigma\circ\pi$.
Thus, a set of BIBCE of $(T,\Theta,\pi,u)$ is robust if and only if the corresponding set of BIBCE of the minimum representation is robust. We use this observation to prove our main result.
\end{remark}

\begin{remark}
In \citet{kajiimorris1997a}, the analyst's model is a complete information game. A correlated equilibrium is defined to be  robust if, for sufficiently small $\varepsilon > 0$, every $\varepsilon$-elaboration has a BNE that is close to the correlated equilibrium. See Section~\ref{KM comparison} for further details.
\end{remark}

\section{Generalized potentials}
\label{Generalized potential}

This section introduces a generalized potential function, which we use to derive a sufficient condition for robustness. 
The function is associated with a special class of communication rules, where each signal received by a player is a subset of the player's actions, interpreted as a recommendation to choose an action from this subset.
We start with discussing such communication rules.

\subsection{$\mathcal{A}$-Decision rule}

As a set of signals for communication rules, 
let $\mathcal{A}_{i}\subset 2^{A_{i}}\backslash\emptyset$
be a covering of $A_{i}$ for each $i\in I$; that is, $\mathcal{A}_{i}$
is a collection of nonempty subsets of $A_{i}$ such that $\bigcup_{X_{i}\in\mathcal{A}_{i}}X_{i}=A_{i}$.
Each signal $X_i \in \mathcal{A}_{i}$ vaguely prescribes an action in $X_i$; that is, it recommends that player $i$ choose an action from $X_{i}$. 
We write
$\mathcal{A}=\{X\mid X=\prod_{i\in I}X_{i},\ X_{i}\in\mathcal{A}_{i}\}$
and ${\mathcal{A}_{-i}}=\{X_{-i}\mid X_{-i}=\prod_{j\neq i}X_{j},\ X_{j}\in{\mathcal{A}_{j}}\}$.

For example, let $A_i=\{0,1\}$ and $\mathcal{\mathcal{{A}}}_{i}=\left\{ \left\{ 0,1\right\} ,\left\{ 1\right\} \right\} $ for each $i\in I$. 
If player $i$ receives $X_i=\{1\}$, the recommendation is to choose action 1. 
If player $i$ receives $X_i=\{0,1\}$, the recommendation allows the player to freely choose an action. This example will be discussed in Section \ref{subsec: Binary-action supermodular game}.

We can describe both a decision rule and a communication rule using a single mapping $\gamma:T\times\Theta\to\Delta(A\times \mathcal{A})$ 
such that $\gamma(a,X|t,\theta)=0$
whenever $a\not\in X$, which assigns a joint probability distribution over $(a,X)\in A\times\mathcal{A}$ to each $(t,\theta)\in T\times\Theta$.  
This mapping is referred to as an $\mathcal{A}$-decision rule, under which player $i$ receives a signal $X_i$ as a vague recommendation and selects an action $a_i$ from $X_i$. 

We say that an $\mathcal{A}$-decision rule $\gamma$ is obedient if the corresponding decision rule in the conjunction is obedient; that is, for each $i\in I$ and $(t_{i},X_{i})\in T_{i}\times \mathcal{A}_{i}$, it holds that 
\begin{equation}
\sum_{a_{-i},\,X_{-i},\,t_{-i},\,\theta}\gamma(a,X|t,\theta)\pi(t,\theta)u_{i}(a,\theta)\geq\sum_{a_{-i},\,X_{-i},\,t_{-i},\,\theta}\gamma(a,X|t,\theta)\pi(t,\theta)u_{i}((a_{i}',a_{-i}),\theta)\label{obedient condition 3}
\end{equation}
for all $a_{i},a_{i}'\in A_{i}$.  
Belief-invariance
of an $\mathcal{A}$-decision rule is defined similarly: $\gamma$ is
belief-invariant if $\gamma(\{a_{i}\}\times A_{-i}\times\{X_{i}\}\times \mathcal{A}_{-i}|t,\theta)$
is independent of $(t_{-i},\theta)$ for all $(a_{i},X_{i})\in A_{i}\times \mathcal{A}_{i}$,
$t_{i}\in T_{i}$, and $i\in I$. 
An obedient belief-invariant $\mathcal{A}$-decision rule $\gamma$ is referred to as a BIBCE, with some abuse of terminology, since the ``decision rule'' component of $\gamma$ is a BIBCE, as demonstrated by the next lemma.

\begin{lemma}\label{BIBCE-C}
If an $\mathcal{A}$-decision rule $\gamma$ is a BIBCE,
then $\sigma\in\Sigma^{BI}$ with $\sigma(a|t,\theta)=\gamma(\{a\}\times \mathcal{A}|t,\theta)$
for all $(a,t,\theta)\in A\times T\times\Theta$ is a BIBCE of $(T,\Theta,\pi,u)$. 
\end{lemma}

\subsection{Generalized potentials and BIBCE}

A generalized potential function \citep{morrisui2005} is defined as a function over $\mathcal{A} \times \Theta$, which contains certain information about players' preferences.

\begin{definition}\label{new def: GP} A bounded function $F:\mathcal{A}\times\Theta\to\mathbb{R}$
is a generalized potential function of $(T,\Theta,\pi,u)$ if, for each $i\in I$ and $P_{i}\in\Delta(A_{-i}\times\mathcal{A}_{-i}\times\Theta)$
such that $P_{i}(A_{-i}\times\mathcal{A}_{-i}\times\Theta^{*})=1$
and $P_{i}(a_{-i},X_{-i},\theta)=0$ whenever $a_{-i}\not\in X_{-i}$,
\begin{equation}
X_{i}\in\arg\max_{X_{i}'\in\mathcal{A}_{i}}\sum_{X_{-i},\theta}P_{i}(X_{-i},\theta)F((X_{i}',X_{-i}),\theta)\label{condition 1 gp}
\end{equation}
implies 
\begin{equation}
X_{i}\cap\arg\max_{a_{i}'\in{A}_{i}}\sum_{a_{-i},\theta}P_{i}(a_{-i},\theta)u_{i}((a_{i}',a_{-i}),\theta)\neq\emptyset,\label{condition 2 gp}
\end{equation}
where $P_{i}(X_{-i},\theta)=\sum_{a_{-i}\in A_{-i}}P_{i}(a_{-i},X_{-i},\theta)$
and $P_{i}(a_{-i},\theta)=\sum_{X_{-i}\in\mathcal{A}_{-i}}P_{i}(a_{-i},X_{-i},\theta)$. 
The value of $F(X,\theta)$ for $\theta\not\in\Theta^{*}$ can be arbitrary. 
\end{definition}

A probability distribution $P_{i}$ is typically derived from an $\mathcal{A}$-decision rule $\gamma$ as player $i$'s
belief over $A_{-i}\times \mathcal{A}_{-i}\times\Theta$ when the opponents' actions and signals follow $\gamma$: 
\[
P_i(a_{-i},X_{-i},\theta)=\sum_{t_{-i}}\sum_{a_i,X_i}\gamma(a,X|t,\theta)\pi(t_{-i},\theta|t_i).
\]
Suppose that player $i$ receives a recommendation $X_{i}\in \mathcal{A}_i$ that satisfies \eqref{condition 1 gp} under this belief; that is, $X_i$ maximizes the expected value of $F$. 
Condition \eqref{condition 2 gp} then requires that at least one action in $X_{i}$ maximizes the expected value of player $i$'s payoff function.

At one extreme, let $\mathcal{A}_{i}=\{A_{i}\}$ for each $i\in I$, where a signal contains no information. 
Clearly, every incomplete information game has a generalized potential with this domain.

At the other extreme, let $\mathcal{A}_{i}=\{\{a_{i}\}\mid a_{i}\in A_{i}\}$ for
each $i\in I$, so that each signal recommends a single action. 
A potential function \citep{monderershapley1996} is a generalized potential function on this domain. 
A function
$v:A\times\Theta\to\mathbb{R}$ is a potential function of $(T,\Theta,\pi,u)$ if 
there exists $q_{i}:A_{-i}\times\Theta\to\mathbb{R}$
such that 
\begin{equation}
u_{i}(a,\theta)=v(a,\theta)+q_{i}(a_{-i},\theta)	
\label{def MS potential 2}
\end{equation}
for all $i\in I$, $a\in A$, and $\theta\in\Theta^*$. 
It is straightforward to show that \eqref{def MS potential 2} implies the condition in Definition \ref{new def: GP} when $F(\{a\},\theta)=v(a,\theta)$.

\begin{lemma}\label{pot and G-pot} 
Suppose that $(T,\Theta,\pi,u)$ has a potential function $v:A\times\Theta\to\mathbb{R}$. Then, $F:\mathcal{A}\times\Theta\to\mathbb{R}$
with $\mathcal{A}_{i}=\{\{a_{i}\}\mid a_{i}\in A_{i}\}$ for each
$i\in I$ and $F(\{a\},\theta)=v(a,\theta)$ for each $(a,\theta)\in A\times\Theta$
is a generalized potential function. 
\end{lemma} 

One of the key properties of a potential function is that if a strategy profile maximizes its expected value, then it constitutes a BNE.
A similar result also holds for BIBCE, even for generalized potential functions.

To see this, fix a generalized potential function $F$ with a domain $\mathcal{A}$. 
Let $\Gamma^{BI}$ denote the set of all 
belief-invariant $\mathcal{A}$-decision rules. 
For each $\gamma\in\Gamma^{BI}$, 
let $\gamma(a|t,\theta)$ and $\gamma(X|t,\theta)$ denote the conditional marginal probabilities of $a$ and $X$, respectively; that is, $\gamma(a|t,\theta)=\sum_{X\in\mathcal{A}}\gamma(a,X|t,\theta)$ and 
$\gamma(X|t,\theta)=\sum_{a\in A}\gamma(a,X|t,\theta)$.

Let $\Gamma^{F}\subset\Gamma^{BI}$ be the set of belief-invariant
$\mathcal{A}$-decision rules that maximize the expected value of a
generalized potential function $F$: 
\[
\Gamma^{F}\equiv
\arg\max_{\gamma\in\Gamma^{BI}}\sum_{X,t,\theta}\gamma(X|t,\theta)\pi(t,\theta)F(X,\theta).
\]
The next lemma shows that $\Gamma^{F}$ contains a BIBCE. 
\begin{lemma}\label{Gmax BIBCE lemma}
There exists a BIBCE $\gamma\in\Gamma^{F}$.
\end{lemma} 

If $\gamma\in\Gamma^{F}$ is a BIBCE, then $\sigma\in\Sigma^{BI}$
with $\sigma(a|t,\theta)=\gamma(a|t,\theta)$ is a BIBCE by Lemma~\ref{BIBCE-C}.
Such a BIBCE is referred to as a GP-maximizing BIBCE. We denote the
set of all GP-maximizing BIBCE by 
\[
\mathcal{E}^{F}\equiv\{\sigma\in\Sigma^{BI}\mid\text{\ensuremath{\gamma\in\Gamma^{F}} is a BIBCE and \ensuremath{\sigma(a|t,\theta)=\gamma(a|t,\theta)}}\}.
\]

\section{Main results}
\label{Main results}

\subsection{Robustness of GP-maximizing BIBCE}

\label{GP-maximizing BIBCE and robustness}

The following main result of this paper shows that $\mathcal{E}^{F}$ is robust.

\begin{theorem}\label{main theorem} If a non-redundant game $(T,\Theta,\pi,u)$
has a generalized potential function $F:\mathcal{A}\times\Theta\to\mathbb{R}$,
then $\mathcal{E}^{F}$ is nonempty and robust. \end{theorem}

Every game admits a generalized potential function with the domain $\mathcal{A}=\{A\}$, in which case $\mathcal{E}^{F}$ is the set of all BIBCE. The theorem implies that if a BIBCE is unique and is the only element of $\mathcal{E}^{F}$, it is robust. 

\begin{corollary}\label{main corollary}
If a non-redundant game $(T,\Theta,\pi,u)$ has a unique BIBCE, then it is robust. 
\end{corollary} 
This result is analogous to the finding in \citet{kajiimorris1997a} that a unique correlated equilibrium of a complete information game is robust.


If a game has a potential function
$v:A\times\Theta\to\mathbb{R}$, it has a generalized potential
function $F:\mathcal{A}\times\Theta\to\mathbb{R}$ with $\mathcal{A}=\{\{a\}\mid a\in A\}$
and $F(\{a\},\theta)=v(a,\theta)$ by Lemma \ref{pot and G-pot}. 
In this case, it is straightforward to show that the set of all GP-maximizing BIBCE coincides with the set of all potential maximizing (P-maximizing, henceforth) belief-invariant decision rules: 
\[
\mathcal{E}^{F}=
\mathcal{E}^{v}=\Sigma^v\equiv\arg\max_{\sigma\in\Sigma^{BI}}\sum_{a,t,\theta}\sigma(a|t,\theta)\pi(t,\theta)v(a,\theta).
\]
Thus, we obtain the following corollary of Theorem \ref{main theorem}.

\begin{corollary}\label{main theorem corollary} 
If a non-redundant game
$(T,\Theta,\pi,u)$ has a potential function $v:A\times\Theta\to\mathbb{R}$,
then $\Sigma^{v}$ is nonempty and robust. 
\end{corollary}

We apply this corollary to the motivating example in the introduction. 
The game has a potential function that is identical to the payoff function in Table \ref{fig1}.
Thus, the expected value of the potential function under the BIBCE in Table \ref{fig3} attains the maximum value of one over all decision rules. Moreover, it can be readily shown that this BIBCE is the unique P-maximizing BIBCE. Consequently, by Corollary \ref{main theorem corollary}, the BIBCE is robust.

\subsection{Proof of Theorem \ref{main theorem}}

\label{Proof of the main theorem}

\label{section: proof} 

It suffices to show that every sequence of $\varepsilon^{k}$-elaborations
$\{(\bar{T}^{k},\bar{\Theta}^{k},\bar{\pi}^{k},\bar{u}^{k})\}_{k=1}^{\infty}$
with $\lim_{k\to\infty}\varepsilon^{k}=0$ has a sequence of BIBCE
$\{\bar{\sigma}^{k}\}_{k=1}^{\infty}$ such that 
\begin{equation}
\lim_{k\to\infty}\inf_{\sigma\in\mathcal{E}^F}\sup_{E\subseteq A\times T^*\times\Theta}\left|\sum_{(a,t,\theta)\in E}\bar{\sigma}^{k}\circ{\bar{\pi}^{k}}(a,(\tau^{k})^{-1}(t),\theta)-\sum_{(a,t,\theta)\in E}\sigma\circ\pi(a,t,\theta)\right|=0,\notag 
\end{equation}
where $\tau^{k}$ is an elaboration mapping 
of $(\bar{T}^{k},\bar{\Theta}^{k},\bar{\pi}^{k},\bar{u}^{k})$. 

We first show that a weaker version of the above condition holds. 
The following lemma focuses on a special sequence of $\varepsilon$-elaborations that share the same sets of types, states, and payoff functions,  differing only in their priors. 
Such a sequence of $\varepsilon$-elaborations is shown to admit a corresponding sequence of BIBCE that converges to some GP-maximizing BIBCE as $\varepsilon$ approaches zero.

\begin{lemma}\label{main theorem lemma} 
Let $(T,\Theta,\pi,u)$
be an incomplete information game with a generalized potential function
$F:\mathcal{A}\times\Theta\to\mathbb{R}$. 
Then, every sequence of $\varepsilon^{k}$-elaborations $\{(T,\Theta,\pi^{k},u)\}_{k=1}^{\infty}$
satisfying $\lim_{k\to\infty}\varepsilon^{k}=0$ and sharing a common elaboration mapping $\tau:T\to T$ has a sequence of
BIBCE $\{\sigma^{k}\}_{k=1}^{\infty}$ such that 
\begin{equation}
\lim_{k\to\infty}\inf_{\sigma\in\mathcal{E}^{F}}\sum_{(a,t,\theta)\in A\times T^*\times\Theta}\left|\sigma^{k}\circ{\pi^{k}}(a,\tau^{-1}(t),\theta)-\sigma\circ\pi(a,t,\theta)\right|=0.\label{key lemma eq}
\end{equation}
\end{lemma}

In the following proof of Theorem \ref{main theorem}, we show that any sequence of $\varepsilon$-elaborations can be transformed into an equivalent special sequence considered in Lemma \ref{main theorem lemma}. 
We then apply Lemma \ref{main theorem lemma} to the transformed sequence, thereby establishing the robustness of GP-maximizing BIBCE.

\begin{proof}[Proof of Theorem \ref{main theorem}]
Let $\{(\bar{T}^{k},\bar{\Theta}^{k},\bar{\pi}^{k},\bar{u}^{k})\}_{k=1}^{\infty}$
be a sequence of $\varepsilon^{k}$-elaborations 
such that $\lim_{k\to\infty}\varepsilon^{k}=0$. 
We regard $T$, $\bar{T}^{k}$, and $\bar{T}^{l}$ as distinct sets  for $k\neq l$, and define the disjoint union $\bar{T}=T\cup\left(\bigcup_{k=1}^{\infty}\bar{T}^{k}\right)$.
We assume that $\bar{\Theta}^{k}\cap\bar{\Theta}^{l}=\Theta^*$, and define $\bar{\Theta}=\bigcup_{k=1}^{\infty}\bar{\Theta}^{k}$.

We construct an equivalent sequence of $\varepsilon^{k}$-elaborations
$\{(\bar{T},\bar{\Theta},\bar{\lambda}^{k},\bar{u})\}_{k=1}^{\infty}$.
Let $\bar{\lambda}^{k}\in\Delta(\bar{T}\times\bar{\Theta})$ be an
extension of $\bar{\pi}^{k}$ to $\bar{T}\times\bar{\Theta}$: $\bar{\lambda}^{k}(\bar{t},\bar{\theta})=\bar{\pi}^{k}(\bar{t},\bar{\theta})$
if $(\bar{t},\bar{\theta})\in\bar{T}^{k}\times\bar{\Theta}^{k}$ and
$\bar{\lambda}^{k}(\bar{t},\bar{\theta})=0$ otherwise. Let $\bar{u}_{i}:A\times\bar{\Theta}\to\mathbb{R}$
be such that $\bar{u}_{i}(\cdot,\theta)=u_{i}(\cdot,\theta)$ if $\theta\in\Theta^{*}$
and $\bar{u}_{i}(\cdot,\bar{\theta})=\bar{u}_{i}^{k}(\cdot,\bar{\theta})$
if $\bar{\theta}\in\bar{\Theta}^{k}\setminus\Theta^{*}$ for each
$i\in I$. Given an elaboration mapping $\tau^{k}$
of $(\bar{T}^{k},\bar{\Theta}^{k},\bar{\pi}^{k},\bar{u}^{k})$, an elaboration mapping $\tau$ of $(\bar{T},\bar{\Theta},\bar{\lambda}^{k},\bar{u})$
is defined as $\tau_{i}(\bar{t}_{i})=\tau_{i}^{k}(\bar{t}_{i})$ if
$\bar{t}_{i}\in\bar{T}_{i}^{k}$ for each $i\in I$. Clearly, $(\bar{T},\bar{\Theta},\bar{\lambda}^{k},\bar{u})$
and $(\bar{T}^{k},\bar{\Theta}^{k},\bar{\pi}^{k},\bar{u}^{k})$ have
the same minimum representation (every player of every type on the
common support has the same belief and the same payoffs in both games),
and $(\bar{T},\bar{\Theta},\bar{\lambda}^{k},\bar{u})$ is also an
$\varepsilon^{k}$-elaboration of $(T,\Theta,\pi,u)$.

We introduce another incomplete information game $(\bar{T},\bar{\Theta},\bar{\pi},\bar{u})$,
where $\bar{\pi}\in\Delta(\bar{T}\times\bar{\Theta})$ is an extension
of $\pi$ to $\bar{T}\times\bar{\Theta}$, i.e., $\bar{\pi}(t,\theta)=\pi(t,\theta)$
if $(t,\theta)\in T\times\Theta$ and $\bar{\pi}(t,\theta)=0$ otherwise.
Note that $(\bar{T},\bar{\Theta},\bar{\pi},\bar{u})$ and $(T,\Theta,\pi,u)$
have the same minimum representation, and $(\bar{T},\bar{\Theta},\bar{\lambda}^{k},\bar{u})$
is an $\varepsilon^{k}$-elaboration of $(\bar{T},\bar{\Theta},\bar{\pi},\bar{u})$.
Because an arbitrary extension of $F$ to $\mathcal{A}\times\bar{\Theta}$
is a generalized potential function of $(\bar{T},\bar{\Theta},\bar{\pi},\bar{u})$,
$\bar{\sigma}:\bar{T}\times\bar{\Theta}\to\Delta(A)$ is a GP-maximizing
BIBCE of $(\bar{T},\bar{\Theta},\bar{\pi},\bar{u})$ if and only if
the restriction of $\bar{\sigma}$ to $T\times\Theta$ is a GP-maximizing
BIBCE of $(T,\Theta,\pi,u)$.

Now, let $\overline{\mathcal{E}}^{F}$ be the set of all GP-maximizing
BIBCE of $(\bar{T},\bar{\Theta},\bar{\pi},\bar{u})$. Then, by Lemma
\ref{main theorem lemma}, $\{(\bar{T},\bar{\Theta},\bar{\lambda}^{k},\bar{u})\}_{k=1}^{\infty}$
has a sequence of BIBCE $\{\bar{\sigma}^{k}\}_{k=1}^{\infty}$ such
that 
\[
\lim_{k\to\infty}\inf_{\bar{\sigma}\in\overline{\mathcal{E}}^{F}}\sum_{(a,t,\theta)\in A\times\bar{T}\times\bar{\Theta}}\left|\bar{\sigma}^{k}\circ{\bar{\lambda}^{k}}(a,\tau^{-1}(t),\theta)-\bar{\sigma}\circ{\bar{\pi}}(a,t,\theta)\right|=0,
\]
which is equivalent to 
\[
\lim_{k\to\infty}\inf_{\bar{\sigma}\in\overline{\mathcal{E}}^{F}}\sup_{E\subseteq A\times\bar{T}\times\bar{\Theta}}\left|\sum_{(a,t,\theta)\in E}\bar{\sigma}^{k}\circ{\bar{\lambda}^{k}}(a,\tau^{-1}(t),\theta)-\sum_{(a,t,\theta)\in E}\bar{\sigma}\circ{\bar{\pi}}(a,t,\theta)\right|=0.
\]
Let $\bar{\xi}^{k}:\bar{T}^{k}\times\bar{\Theta}^{k}\to\Delta(A)$
be the restriction of $\bar{\sigma}^{k}:\bar{T}\times\bar{\Theta}\to\Delta(A)$
to $\bar{T}^{k}\times\bar{\Theta}^{k}$. Then, $\bar{\xi}^{k}$ is
a BIBCE of $(\bar{T}^{k},\bar{\Theta}^{k},\bar{\pi}^{k},\bar{u}^{k})$,
and $\{\bar{\xi}^{k}\}_{k=1}^{\infty}$ satisfies
\[
\lim_{k\to\infty}\inf_{\sigma\in\mathcal{E}^{F}}\sup_{E\subseteq A\times T\times\Theta}\left|\sum_{(a,t,\theta)\in E}\bar{\xi}^{k}\circ{\bar{\pi}^{k}}(a,(\tau^{k})^{-1}(t),\theta)-\sum_{(a,t,\theta)\in E}\sigma\circ\pi(a,t,\theta)\right|=0
\]
because $\{\bar{\sigma}\circ{\bar{\pi}}\in\Delta(A\times\bar{T}\times\bar{\Theta})\mid\bar{\sigma}\in\overline{\mathcal{E}}^{F}\}$
coincides with $\{\sigma\circ{\pi}\in\Delta(A\times T\times\Theta)\mid\sigma\in{\mathcal{E}}^{F}\}$
on their common support. Therefore, $\mathcal{E}^{F}$ is robust. 
\end{proof}

\subsection{Proof outline for Lemma \ref{main theorem lemma}}

We sketch the proof of Lemma \ref{main theorem lemma}, focusing on the special case where $(T,\Theta,\pi,u)$ has a potential function~$v$ and a P-maximizing BIBCE is unique, i.e., $\calE^v=\Sigma^v=\{\sigma^*\}$.
 We adopt a potential function satisfying the following condition. 
Let $\phi:\Theta\to\Theta^*$ be a mapping such that the $i$-th component of $\phi(\theta)=(\phi_i(\theta))_{i\in I}$ equals $\theta_i$ if $\theta_i\in\Theta_i^*$. 
Note that $\phi$ is the identity mapping when restricted to $\Theta^*$. 
Assume that $v$ satisfies 
\begin{equation}
v(a,\theta)=v(a,\phi(\theta)) \text{ for all }(a,\theta)\in A\times\Theta, \label{choice of potential} 	
\end{equation}
which is possible because
the definition of potential functions does not impose any conditions on their values when $\theta\not\in \Theta^*$. 
Then, for player~$i$ with $\theta_i\in\Theta_i^*$,
the best response under $u_i(a,\theta)$ coincides with that under $v(a,\theta)$ by \eqref{def MS potential 2} and~\eqref{choice of potential}.\footnote{Note that $u_i(a,\theta)=u_i(a,\phi(\theta))=v(a,\phi(\theta))+q_i(a_{-i},\phi(\theta))
=v(a,\theta)+q_i(a_{-i},\phi(\theta))$.}

The proof proceeds in three steps. 
In the first step, we construct a
candidate for a sequence of BIBCE $\{\sigma^{k}\}_{k=1}^{\infty}$ satisfying \eqref{key lemma eq}. 
In so doing, let
$T_{i}^{k}\equiv\{t_{i}\in T_{i}\mid\pi^{k}(\Theta_{i}^{*}\times\Theta_{-i}|t_{i})=1\}$
denote the set of player $i$'s types in an $\varepsilon^k$-elaboration $(T,\Theta,\pi^{k},u)$ who believe that their payoff functions
are the same as those in the analyst's model $(T,\Theta,\pi,u)$. 
Note that $\pi^{k}(T^{k})\geq1-\varepsilon^{k}$
by the second condition of Definition \ref{def: 2}.
We call types in $T^k_i$ standard types, and all other types perturbed types. 
For a type profile $t\in T$, we denote the set of players with standard types by $S^{k}(t)\equiv\{i\in I\mid t_{i}\in T_{i}^{k}\}$.

Consider a decision rule $\sigma^k$ of $(T,\Theta,\pi^{k},u)$, where standard types and perturbed types choose actions independently of each other, conditional on $(t,\theta)$; that is, 
\begin{equation}
	\sigma^k(a|t,\theta)=
\sigma_{S^k(t)}^k(a_{S^k(t)}|t,\theta)\cdot \sigma_{-S^k(t)}^k(a_{-S^k(t)}|t,\theta),
\label{candidate condition 1}
\end{equation}
where $\sigma_{S}(a_{S}|t,\theta)\equiv\sigma(\{a_{S}\}\times A_{-S}|t,\theta)$ and $\sigma_{-S}(a_{-S}|t,\theta)\equiv\sigma(\{a_{-S}\}\times A_{S}|t,\theta)$. 
Let $\Sigma^k$ denote the set of all belief-invariant decision rules of $(T,\Theta,\pi^{k},u)$ that satisfy \eqref{candidate condition 1}.
For each $\sigma^k\in \Sigma^k$, we refer to $\sigma_{S^k(t)}$ and $\sigma_{-S^k(t)}$ as the standard types' rule and the perturbed types' rule, respectively.

Using the Kakutani--Fan--Glicksberg fixed point theorem, we can show the existence of a BIBCE $\sigma^k\in \Sigma^k$ satisfying the following condition: The standard types' rule $\sigma^k_{S^k(t)}$ maximizes the expected value of a potential function: 
\begin{align}
\sum_{a,t,\theta}\sigma^k\circ\pi^{k}(a,t,\theta)v(a,\theta)\geq 
\sum_{a,t,\theta}\sigma'\circ\pi^{k}(a,t,\theta)v(a,\theta)
\label{potential maximize requirement}
\end{align}
for any $\sigma'\in \Sigma^k$ such that the perturbed types' rule is the same as that of $\sigma^k$. 

\begin{claim}
For each $k$, $(T,\Theta,\pi^k,u)$ has a BIBCE $\sigma^k\in \Sigma^k$ that satisfies \eqref{potential maximize requirement} for any $\sigma'\in \Sigma^k$ with $\sigma^k_{-S^k(t)}=\sigma'_{-S^k(t)}$. 
\end{claim}

Consider the sequence of BIBCE $\{\sigma^{k}\}_{k=1}^{\infty}$. 
In the second step, we show that the limit of the left-hand side of~\eqref{potential maximize requirement} as $k\to\infty$ is at least as large as that under the potential-maximizing BIBCE $\sigma^*$ of the analyst's model.
\begin{claim}
The sequence of BIBCE $\{\sigma^{k}\}_{k=1}^{\infty}$ satisfies
\begin{equation}
\liminf_{k\to \infty}\sum_{a,t,\theta}\sigma^{k}\circ\pi^{k}(a,t,\theta)v(a,\theta)\geq\sum_{a,t,\theta}{\sigma}^{*}\circ\pi(a,t,\theta)v(a,\theta).\label{final key eq 1''''}
\end{equation}
\end{claim}

To see why, 
let $\hat \sigma^{k}\in\Sigma^{k}$ be the decision rule in which perturbed types follow $\sigma^k$ and standard types follow $\sigma^*$.\footnote{In particular, for standard types, $
\hat{\sigma}_{S^{k}(t)}^{k}(a_{S^{k}(t)}|t,\theta)=\sigma^*_{S^{k}(t)}(a_{S^{k}(t)}|\tau(t),\theta)$.} Then, by~\eqref{potential maximize requirement}, we obtain 
\begin{equation}
\sum_{a,t,\theta}\sigma^{k}\circ\pi^{k}(a,t,\theta)v(a,\theta)\geq\sum_{a,t,\theta}\hat{\sigma}^{k}\circ\pi^{k}(a,t,\theta)v(a,\theta).\notag
\end{equation}
The right-hand side of the above inequality converges to that of \eqref{final key eq 1''''} since $\lim_{k\to \infty}\pi^k(T^k)=1$.

In the final step, we show that 
 $\{\sigma^k\}_{k=1}^\infty$ satisfies \eqref{key lemma eq}. 
We can rewrite \eqref{final key eq 1''''} as 
\begin{equation}
\liminf_{k\to \infty}\sum_{a,t,\theta}\sigma^{k}\circ{\pi^{k}}(a,\tau^{-1}(t),\theta)v(a,\theta)\geq\sum_{a,t,\theta}{\sigma}^{*}\circ\pi(a,t,\theta)v(a,\theta),\label{final key eq 1eta}
\end{equation}
since $\{\tau^{-1}(t)\}_{t\in T^*}$ partitions $T$. This inequality is shown to imply the following, which in turn implies~\eqref{key lemma eq}.

\begin{claim}
It holds that 
\begin{equation}
\lim_{k\to\infty}\sum_{a,t,\theta}\left|\sigma^{k}\circ{\pi^{k}}(a,\tau^{-1}(t),\theta)-\sigma^*\circ\pi(a,t,\theta)\right|=0.\label{key lemma eqeta}
\end{equation}
\end{claim}

To establish this, let $\eta^{k}\in\Delta(A\times T\times\Theta)$
be given by $\eta^{k}(a,t,\theta)\equiv\sigma^{k}\circ{\pi^{k}}(a,\tau^{-1}(t),\theta)$. 
The sequence $\{\eta^{k}\}_{k=1}^{\infty}$ is tight, so it has a convergent subsequence with a limit point $\eta^{\star}$. 
Then, we can construct a belief-invariant decision rule $\sigma^\star$ satisfying $\sigma^\star(a|t,\theta)=\eta^\star(a,t,\theta)/\pi(t,\theta)$ for all $(t,\theta)$ with $\pi(t,\theta)>0$. 
By~\eqref{final key eq 1eta}, it holds that
$\sum_{a,t,\theta}{\sigma}^\star\circ\pi(a,t,\theta)v(a,\theta)
\geq\sum_{a,t,\theta}{\sigma}^{*}\circ\pi(a,t,\theta)v(a,\theta)$. 
Since $\sigma^*$ is the unique P-maximizing belief-invariant decision rule, we must have 
${\eta}^\star={\sigma}^\star\circ\pi
={\sigma}^{*}\circ\pi$, which holds for any convergent subsequence of $\{\eta^{k}\}_{k=1}^{\infty}$.
This implies that $\{\eta^{k}\}_{k=1}^{\infty}$ converges to ${\sigma}^{*}\circ\pi$ and \eqref{key lemma eqeta} holds.

\section{Supermodular games}

\label{Robust BIBCE of supermodular games}

\subsection{Supermodular potential games}

The robust BIBCE in the motivating example is not a BNE. 
This raises a natural question: in what class of games can a BNE be robust?
We show that the robust set of BIBCE given by Theorem \ref{main theorem} contains BNE if the game is a supermodular potential game. 
In particular, if the robust set is a singleton, then the robust BIBCE is a BNE.

Let $A_{i}$ be linearly ordered with $\geq_{i}$. We write $\geq_I$
for the product order: for $a,b\in A$, $a\geq_Ib$ if and only if $a_{i}\geq_{i}b_{i}$ for
all $i\in I$. We say that $(T,\Theta,\pi,u)$ is a supermodular game if, for each $\theta\in\Theta$, the ex-post game
with a payoff function profile $(u_{i}(\cdot,\theta))_{i\in I}$ exhibits 
strategic complementarities; that is, for each $i\in I$ and $a,b\in A$
with $a\geq_I b$, 
\[
u_{i}((a_{i},a_{-i}),\theta)-u_{i}((b_{i},a_{-i}),\theta)\geq u_{i}((a_{i},b_{-i}),\theta)-u_{i}((b_{i},b_{-i}),\theta).
\]

For each belief-invariant decision rule $\sigma\in\Sigma^{BI}$, we define a pure strategy profile $\overline{\sigma}=(\overline{\sigma}_{i})_{i\in I}$, where for each $i\in I$ and $t_{i}\in T_{i}$,  player $i$ of type $t_i$ chooses the maximum action in the support of $\sigma_{i}(\cdot|t_{i})$; that is, $\bar \sigma_i(\cdot|t_i)$ places probability one on $\max\{a_{i}\mid \sigma_i(a_i|t_i)=\sigma\left(\{a_{i}\}\times A_{-i}|\left(t_{i},t_{-i}\right),\theta\right)>0\}$.  
Similarly, we define another pure strategy profile $\underline{\sigma}=(\underline{\sigma}_{i})_{i\in I}$, where player $i$ of type $t_i$ chooses the minimum action in the support of $\sigma_{i}(\cdot|t_{i})$.

The following proposition establishes that if $\sigma$ is a P-maximizing BIBCE, then 
the strategy profiles $\overline{\sigma}$ and $\underline{\sigma}$ are also P-maximizing BIBCE. 

\begin{proposition}\label{supermodular result} 
Assume that a supermodular
game $(T,\Theta,\pi,u)$ has a potential function $v$. Then, for each P-maximizing BIBCE $\sigma\in\calE^{v}$, it holds that $\overline{\sigma},\underline{\sigma}\in\calE^{v}$. 
\end{proposition}

This proposition implies that $\overline{\sigma}$ and $\underline{\sigma}$ are BNE that maximize the expected value of the potential function (P-maximizing BNE).
Thus, if a P-maximizing BNE is unique, then 
we must have $\sigma=\overline{\sigma}=\underline{\sigma}$, which is a robust BNE.

As an example of such a BNE, we consider
a binary-action supermodular game with two players, 
each choosing either action 1 or action 0 (which can be interpreted as ``invest'' and ``not invest,'' respectively). 
The set of states is given by $\Theta=\{0,1,2,3,\ldots\}$. 
A payoff table and a potential function for each state $n\in\Theta$ are given below, where $0<r<1$.
\[
\begin{tabular}{l|cc}
\multicolumn{3}{c}{A payoff table} \\
  &  \text{action 1}  &  \text{action 0} \\
\hline  \text{action 1}  &  \ensuremath{{r}^{n+1},{r}^{n+1}}  &  \ensuremath{{r}^{n+1}-1,0} \\
\text{action 0}  &  \ensuremath{0,{r}^{n+1}-1}  &  \ensuremath{0,0} 
\end{tabular}
\qquad
\begin{tabular}{l|cc}
\multicolumn{3}{c}{A potential function} \\
  &  \text{action 1}  &  \text{action 0} \\
\hline  \text{action 1}  &  \ensuremath{{r}^{n+1}}  &  \ensuremath{0} \\
\text{action 0}  &  \ensuremath{0}  &  \ensuremath{1-{r}^{n+1}} 
\end{tabular}
\]

State $n\in\Theta$ occurs with probability $p(1-p)^{n}>0$,
where $p\in(0,1)$. 
Each player has the following information partition of $\Theta$: 
player $1$'s partition is
$\{\{0\},\{1,2\},\{3,4\},\ldots\}$, and player $2$'s 
partition is $\{\{0,1\},\{2,3\},\{4,5\},\ldots\}$, which are analogous to those in the email game. 
The type of a player with a partition $\{k-1,k\}$ (or $\{0\}$) is referred
to as type $k$ (or type $0$). Thus, the sets of player 1's types and player 2's
types are $T_{1}=\{0,2,4,\ldots\}$ and $T_{2}=\{1,3,5,\ldots\}$,
respectively. 
This game can be interpreted as a global game \citep{carlssonvandamme1993a} with discretized signals and no dominance regions.

We represent a pure-strategy profile as a sequence of actions $x=(x_{k})_{k=0}^{\infty}\in\{0,1\}^{T_{1}\cup T_{2}}$, where $x_{k}$ is the action of type $k$. 
Let $s^{\tau}=(s_k^\tau)_{k=0}^\infty$ denote 
a monotone strategy profile: 
\[
s_{k}^{\tau}
=\begin{cases}
	1 & \text{if $k\leq \tau-1$},\\
	0 & \text{if $k\geq \tau$},
\end{cases}
\]
where $\tau$ is a non-negative integer or infinite ($\tau=\infty$). 
If $r$ is sufficiently close to one, 
this game has exactly three pure-strategy BNE: $s^0$, $s^\infty$, and $s^{\tau^*}$, where $\tau^*\geq 2$ is the smallest integer satisfying ${r}^{\tau^*}<(1-p)/(1+r(1-p))$, provided no integer $\tau$ satisfies this condition with equality.\footnote{This condition ensures that type $\tau^*$'s best response is action 0 when type $\tau^*-1$ chooses action 1 and type $\tau^*+1$ chooses action 0. The conditional expected value of the potential function for type $\tau^*$ is $r^{\tau^*}/(2-p)$ under action 1 and $(1-p)(1-r^{\tau^*+1})/(2-p)$ under action 0. If the latter is greater, action 0 is the best response.} 


The expected value of the potential function under a strategy profile $x$ is calculated as 
\[
f(x)\equiv \sum_{k=0}^{\infty}p(1-p)^{k}\Big({r}^{k+1}x_{k}x_{k+1}+(1-r^{k+1})(1-x_{k})(1-x_{k+1})\Big).
\]
Using this formula, we can verify that $f(s^{\tau^*})>f(s^0)$ and $f(s^{\tau^*})>f(s^\infty)$. 
Thus, $s^{\tau^*}$ is the unique P-maximizing BNE, which is a robust BNE by Proposition
\ref{supermodular result}.

\subsection{Binary-action supermodular games}\label{subsec: Binary-action supermodular game}

We consider a binary-action supermodular game without a potential function. 
The complete-information case has been extensively studied in the robustness 
literature \citep{oyamatakahashi2020,morrisetal2020}. 

Let $(T,\Theta,\pi,u)$ be a supermodular game with $A_{i}=\{0,1\}$ for each $i\in I$. 
We denote by $\mathbf{1}$ the action profile in which all players choose action $1$, 
and by $\sigma^{\mathbf{1}}$ the decision rule in which every type of every player 
chooses action $1$. 

Suppose that there exist a function $v:A\times\Theta\to\mathbb{R}$ 
with $
v(\mathbf{1},\theta)>v(a,\theta)$  for all $a\neq \mathbf{1}$  and $\theta\in \Theta$
and a constant $\lambda_{i}>0$ for each $i\in I$ that satisfy
\begin{equation}
\lambda_{i}(u_{i}((1,a_{-i}),\theta)-u_{i}((0,a_{-i}),\theta))\geq v((1,a_{-i}),\theta)-v((0,a_{-i}),\theta)\label{LP condition}
\end{equation}
for all $a_{-i}\in A_{-i}$ and $\theta\in\Theta$. 
If $\lambda_i=1$ for all $i\in I$ and \eqref{LP condition} holds with equality, 
then $v$ is a potential function. 
By construction, $\sigma^{\mathbf{1}}$ is a BNE for any prior $\pi$, since 
\begin{equation}
u_{i}((1,\mathbf{1}_{-i}),\theta)-u_{i}((0,\mathbf{1}_{-i}),\theta)\geq \left(v((1,\mathbf{1}_{-i}),\theta)-v((0,\mathbf{1}_{-i}),\theta)\right)/\lambda_{i}>0 \notag
\end{equation}
for all $\theta\in\Theta$.
\citet{morrisui2005} introduce this function for the complete-information case, 
where $\Theta$ is a singleton, and call it a monotone potential 
function.\footnote{See Definition 11 of \citet{morrisui2005}.} 
We use the same term for the incomplete information case.

The next lemma, which follows directly from \citet{morrisui2005}, constructs a generalized potential function from $v$.

\begin{lemma}
If a binary-action supermodular game has a monotone potential function $v$, 
it has a generalized potential function $F:\calA\times\Theta\to\mathbb{R}$
with $\calA_{i}=\{\{0,1\},\{1\}\}$ for each $i\in I$, given by 
\begin{equation}
F(X,\theta)	= v((\min X_i)_{i\in I},\theta)
\ \text{ for all $(X,\theta)\in \calA\times\Theta$.}
\end{equation}
\end{lemma}

Since $\sigma^{\mathbf{1}}$ is the unique GP-maximizing BIBCE, it is robust under every prior $\pi$ by Theorem~\ref{main theorem}.
When $\Theta$ is a singleton, the robustness of $\sigma^{\mathbf{1}}$ implies the 
robustness of the action profile $\mathbf{1}$ in the sense of \citet{kajiimorris1997a} 
(see Section~\ref{KM comparison}), which also follows from \citet{morrisui2005}. 
\citet{oyamatakahashi2020} establish a generic converse: for a generic binary-action 
supermodular game, if there is no monotone potential function, then, for every 
$\varepsilon>0$, there exists an $\varepsilon$-elaboration in which ``all $0$'' 
survives the iterated elimination of strictly dominated strategies.
Thus, generically, the robustness of $\mathbf{1}$ requires the existence of a monotone potential function.

Whether an analogous converse holds under incomplete information is an open question. Suppose that $\mathbf{1}$ is robust under every prior $\pi$. Then, for each $(t,\theta)\in T\times\Theta$, $\mathbf{1}$ is robust under the degenerate prior concentrated on $(t,\theta)$, which induces a complete-information game. Thus, the result of \citet{oyamatakahashi2020} suggests that, generically, the complete-information game associated with each $\theta$ admits a monotone potential function $v(\cdot,\theta)$ satisfying the statewise counterpart of \eqref{LP condition}, with a positive constant $\lambda_i(\theta)$ for each~$i$. However, $\lambda_i(\theta)$ may vary with $\theta$, whereas \eqref{LP condition} requires $\lambda_i$ to be constant across $\theta\in\Theta$. 

\section{Discussion}\label{discussion section}

We discuss the remaining issues regarding the relationship between the robustness analysis in the complete information setting and our analysis in the incomplete information setting.

\subsection{$\varepsilon$-elaborations in \citet{kajiimorris1997a}}\label{KM comparison}

The framework of \citet{kajiimorris1997a} is summarized as follows. 
The analyst's model is a complete information game, where $T=T^{*}=\{t\}$ and $\Theta=\Theta^{*}=\{\theta\}$ are singletons. 
An $\varepsilon$-elaboration is defined as an incomplete information game that satisfies the first two conditions of Definition~\ref{def: 2}. 
The solution concept for the analyst's model is a correlated equilibrium, and that for $\varepsilon$-elaborations is a BNE.

The set of $\varepsilon$-elaborations defined in \citet{kajiimorris1997a} includes the set of $\varepsilon$-elaborations defined in our paper in the case of complete information games. However, our definition is equivalent to theirs in the following sense.

\begin{lemma} \label{KM equivalence}
Suppose that $T=T^{*}=\{t\}$ and $\Theta=\Theta^{*}=\{\theta\}$.
If $(\bar{T},\bar{\Theta},\bar{\pi},\bar{u})$ satisfies the first two conditions of Definition \ref{def: 2} for $\varepsilon>0$, then it is a $\sqrt{\varepsilon}$-elaboration.
\end{lemma} 

Thus, the key distinction between our robustness concept and that of \citet{kajiimorris1997a} lies in the equilibrium concept used for $\varepsilon$-elaborations: we adopt BIBCE, whereas \citet{kajiimorris1997a} adopt BNE.
\citet{pram2019} proposes a weaker definition of robust equilibria in complete information games than that of \citet{kajiimorris1997a} by adopting agent-normal form correlated equilibria as the solution concept for $\varepsilon$-elaborations.
Since both BNE and agent-normal form correlated equilibria are special cases of BIBCE, our definition of robustness, when applied to complete information games, is weaker than those of \citet{pram2019} and \citet{kajiimorris1997a}.

\subsection{The critical path theorem}

\citet{kajiimorris1997a} study common belief events and derive a lower bound on their probability. 
This result, known as the critical path theorem, is used to establish a sufficient condition for the robustness. \citet{oyamatakahashi2020} extend the theorem and prove its generic converse, thus establishing a fundamental link between robustness and monotone potentials as discussed in Section \ref{subsec: Binary-action supermodular game}.

Given the significance of the critical path theorem, it is natural to examine its counterpart in the incomplete information setting.
To sketch this, consider a binary-action supermodular game $(T,\Theta,\pi,u)$. 
For each $i\in I$, fix a subset of types $E_i\subset T_i$. 
Construct an associated fictitious game by preserving the belief and payoff function for player $i$ of type $t_i \in E_i$, while forcing types $t_i \notin E_i$ to choose action 0 as their dominant action. This fictitious game then constitutes an $\varepsilon$-elaboration with $\varepsilon \equiv 1 - \pi(E)$, where $E=\prod_{i\in I}E_i$.

Consider the largest BNE in the fictitious game, which is the largest pure strategy profile that survives the iterated elimination of strictly dominated strategies. 
Let $CB_i^u(E)$ denote the set of player $i$'s types who choose action 1 in the largest BNE and define $CB^u(E)=\prod_{i\in I}CB_i^u(E) \subset E\subset T$. 
This set $CB^u(E)$ can be interpreted as a common belief event in the sense that players have a common belief that action 1 is a best response in the fictitious game when $t\in CB^u(E)$ is realized.

The critical path theorem provides a lower bound for the probability of $CB^u(E)$ in terms of $\varepsilon$. 
We can establish the following lower bound.

\begin{lemma}\label{critical lemma}
If $(T,\Theta,\pi,u)$ admits a monotone potential function $v$, then
\begin{equation}
\pi(CB^{u}(E))\geq1-\kappa(1-\pi(E)), \text{ where }
\kappa\equiv1+\frac{\sup_{S\subseteq S'\neq I,\theta\in\Theta}v(\mathbf{1}_{S},\theta)-v(\mathbf{1}_{S'},\theta)}{\inf_{S\neq I,\theta\in\Theta}v(\mathbf{1},\theta)-v(\mathbf{1}_{S},\theta)}>0.\label{critical path claim 1.1}
\end{equation}
\end{lemma}
By \eqref{critical path claim 1.1}, $\pi(CB^{u}(E))$ converges to one as $\varepsilon\to 0$.
This implies that if $\varepsilon$ is sufficiently small, the $\varepsilon$-elaboration has a BNE in which every player chooses action 1 with probability close to one.
\citet{oyamatakahashi2020} establish \eqref{critical path claim 1.1} in the special case where $u$ and $v$ are independent of $\theta$.

\appendix

\begin{center}
\Large{{\bf Appendix}}
\end{center}

\setcounter{theorem}{0} \setcounter{lemma}{0} \setcounter{proposition}{0}
\setcounter{definition}{0} 
\numberwithin{equation}{section} \global\long\def\thetheorem{\Alph{theorem}}%
\global\long\def\thelemma{\Alph{lemma}}%
\global\long\def\theproposition{\Alph{proposition}}%
\global\long\def\thedefinition{\Alph{definition}}%

This appendix contains omitted proofs. Throughout these proofs, for any countable set $S$,
we regard the set of probability distributions $\Delta(S)$ as a subset
of the linear space $\{f:S\to\mathbb{R}\mid\sum_{s\in S}|f(s)|<\infty\}$
endowed with the $l_{1}$-norm $\|f\|_{1}=\sum_{s\in S}|f(s)|$. Because
$S$ is countable, it is straightforward to show that the topology
of weak convergence in $\Delta(S)$ coincides with the topology induced
by the $l_{1}$-norm in $\Delta(S)$. Thus, the following result holds by Prohorov's theorem \citep[see][]{billingsley1999}, which will be used throughout.

\begin{lemma}\label{prohorov}
Let $\Delta(S)$ be endowed with the topology induced by the $l_{1}$-norm.
If $P\subset\Delta(S)$ is tight, i.e., for any $\varepsilon>0$,
there exists a finite set $K^{\varepsilon}\subset S$ such that $p(K^{\varepsilon})>1-\varepsilon$
for all $p\in P$, then the closure of $P$ is compact. Conversely,
if the closure of $P\subset\Delta(S)$ is compact, then $P$ is tight.
\end{lemma}

\section{Proofs for Section \ref{Elaborations of incomplete information games}}

\subsection{Proof of Lemma \ref{lemma equivalence elaboration}}

To show the ``if'' part, suppose that, for each $i\in I$, there
exists a mapping $\tau_{i}:\bar{T}_{i}\to T_{i}^*$ satisfying \eqref{lemma 1:1}
and \eqref{lemma 1:2}. Let $M_{i}=\bar{T}_{i}$ and $\mu_{i}:\bar{T}_{i}\to M_{i}$
be such that $\mu_{i}(\bar{t}_{i})=\bar{t}_{i}$ for all $\bar{t}_{i}\in\bar{T}_{i}$,
by which the mapping $\bar{t}_{i}\mapsto(\tau_{i}(\bar{t}_{i}),\mu_{i}(\bar{t}_{i}))$
is one-to-one. Consider a communication rule $\rho:T\times\Theta\to\Delta(M)$
such that $\rho(\bar{t}|t,\theta)\pi(t,\theta)=\bar{\pi}(\bar{t},\theta)$
if $\tau(\bar{t})=t$ and $\rho(\bar{t}|t,\theta)=0$ otherwise. 
When $\pi(t,\theta)=0$ and $\tau(\bar{t})=t$, choose 
$\rho (\bar t|t,\theta)=\prod_i{\bar{\pi}(\bar{t}_{i})}/{\pi(t_{i})}$. Such $\rho$ exists by \eqref{lemma 1:1} and satisfies  \eqref{def:elabo}.   
Then, for $(t,\theta)\in T^{*}\times\Theta^{*}$ with $\pi(t,\theta)>0$
and $\bar{t}_{i}\in\bar{T}_{i}^{*}$ with $t_{i}=\tau_{i}(\bar{t}_{i})$,
\begin{align*}
\rho &(\{\bar{t}_{i}\}\times\bar{T}_{-i}|t,\theta)
=\frac{\bar{\pi}(\{\bar{t}_{i}\}\times\tau_{-i}^{-1}(t_{-i}),\theta)}{\pi(t,\theta)}=\frac{\bar{\pi}(\tau_{-i}^{-1}(t_{-i}),\theta|\bar{t}_{i})\times\bar{\pi}(\bar{t}_{i})}{\pi(t_{-i},\theta|t_{i})\times\pi(t_{i})}=\frac{\bar{\pi}(\bar{t}_{i})}{\pi(t_{i})}
\end{align*}
by \eqref{lemma 1:2}. Because $\rho(\{\bar{t}_{i}\}\times\bar{T}_{-i}|t,\theta)=\bar{\pi}(\bar{t}_{i})/\pi(t_{i})$
is independent of $t_{-i}$ and $\theta$, $\rho$ is a belief-invariant
communication rule, and thus $(\bar{T},\Theta,\bar{\pi},u)$ is an
elaboration of $(T,\Theta,\pi,u)$.

To show the ``only if'' part, suppose that $(\bar{T},\Theta,\bar{\pi},u)$
is an elaboration of $(T,\Theta,\pi,u)$; that is, there exists a
belief-invariant communication rule $\rho$ and mappings $\tau_{i}:\bar{T}_{i}\to T_{i}$
and $\mu_{i}:\bar{T}_{i}\to M_{i}$ for each $i\in I$ satisfying
the condition in Definition \ref{def: elaboration}. We show that $\tau$ and $\bar{\pi}$ satisfy \eqref{lemma 1:1} and
\eqref{lemma 1:2}. Let $(t,\theta)\in T^{*}\times\Theta^{*}$ and
$\bar{t}_{i}\in\bar{T}_{i}^*$ be such that $\tau_{i}(\bar{t}_{i})=t_{i}$.
By \eqref{def:elabo}, 
\[
\bar{\pi}(\tau^{-1}(t),\theta)=\pi(t,\theta)\sum_{\bar{t}:\tau(\bar{t})=t}\rho(\mu(\bar{t})|t,\theta)=\pi(t,\theta),
\]
so \eqref{lemma 1:1} holds. Moreover, by \eqref{def:elabo} again,
\begin{align*}
\bar{\pi}(\tau_{-i}^{-1}(t_{-i}),\theta|\bar{t}_{i}) 
 & =\frac{\pi(t,\theta)\left(\sum_{\bar{t}_{-i}:\tau_{-i}(\bar{t}_{-i})=t_{-i}}\rho(\mu(\bar{t})|t,\theta)\right)}{\sum_{t_{-i}',\theta'}\pi((t_{i},t_{-i}'),\theta')\left(\sum_{\bar{t}_{-i}:\tau_{-i}(\bar{t}_{-i})=t_{-i}'}\rho(\mu(\bar{t})|(t_{i},t_{-i}'),\theta')\right)} 
 =\pi(t_{-i},\theta|t_{i}),
\end{align*}
where the last equality holds because $\sum_{\bar{t}_{-i}:\tau_{-i}(\bar{t}_{-i})=t_{-i}}\rho(\mu(\bar{t})|t,\theta)$
is independent of $t_{-i}$ and $\theta$ by the belief-invariance
of $\rho$. Thus, \eqref{lemma 1:2} holds. 
\hfill $\square$

\section{Proofs for Section \ref{Generalized potential}}

For each $\gamma\in\Gamma^{BI}$, 
the conditional distribution of $(a_i,X_i)$ given $t_i$ is denoted by $\gamma_{i}(a_{i},X_{i}|t_{i})$. 
Belief-invariance implies that 
$\gamma_{i}(a_{i},X_{i}|t_{i})=\gamma(\{a_{i}\}\times A_{-i}\times\{X_{i}\}\times \mathcal{A}_{-i}|t,\theta)$, which will be used in the proofs.  
We write 
$\gamma_{i}(a_{i}|t_{i})=\sum_{X_{i}\in\mathcal{A}_{i}}\gamma_{i}(a_{i},X_{i}|t_{i})$ 
and $\gamma_{i}(X_{i}|t_{i})=\sum_{a_{i}\in{A}_{i}}\gamma_{i}(a_{i},X_{i}|t_{i})$.

\subsection{Proof of Lemma \ref{BIBCE-C}}

Let an $\mathcal{A}$-decision rule $\gamma$ be a BIBCE. Because $\gamma$ is obedient, \eqref{obedient condition 3} holds. By taking the summation of each side of \eqref{obedient condition 3} over $X_{i}\in \mathcal{A}_{i}$, we obtain \eqref{obedient condition 2},
and thus $\sigma$ is obedient. In addition, because $\gamma$ is belief-invariant,
$\gamma(\{a_{i}\}\times A_{-i}\times\{X_{i}\}\times \mathcal{A}_{-i}|t,\theta)=\gamma_{i}(a_{i},X_{i}|t_{i})$,
which implies $\sigma(\{a_{i}\}\times A_{-i}|t,\theta)=\sum_{X_{i}\in \mathcal{A}_{i}}\gamma_{i}(a_{i},X_{i}|t_{i})$.
Thus, $\sigma$ is belief-invariant as well.~\hfill $\square$

\subsection{Proof of Lemma \ref{Gmax BIBCE lemma}}

Each $\gamma\in\Gamma^{BI}$, together
with a prior $\pi$, defines a joint probability distribution $\gamma\circ\pi$ over $A\times\mathcal{A}\times T\times\Theta$ 
given by $\gamma\circ\pi(a,X,t,\theta)=\gamma(a,X|t,\theta)\pi(t,\theta)$,
which is referred to as a distributional (belief-invariant) $\mathcal{A}$-decision rule. 
We denote the set of all such rules by 
\[
\Gamma^{BI}\circ\pi=\{\gamma\circ\pi\in\Delta(A\times\mathcal{A}\times T\times\Theta)\mid\gamma\in\Gamma^{BI}\}
\]
and identify it with $\Gamma^{BI}$; that is, 
we regard $\Gamma^{BI}$ as the set of equivalence classes induced
by each $\gamma\circ{\pi}\in\Gamma^{BI}\circ{\pi}$, where
$\gamma$ and $\gamma'$ are equivalent if $\gamma\circ\pi=\gamma'\circ\pi$.
It can be verified that $\Gamma^{BI}\circ{\pi}$ is a tight closed subset of
$\Delta(A\times\mathcal{A}\times T\times\Theta)$. Thus, by Lemma
\ref{prohorov}, it is compact. 
Therefore, 
\[
\Gamma^{F}\circ\pi\equiv\arg\max_{\gamma\circ\pi\in\Gamma^{BI}\circ\pi}\sum_{a,X,t,\theta}\gamma\circ\pi(a,X,t,\theta)F(X,\theta)
\]
is nonempty, implying that $\Gamma^{F}$ is also nonempty. 

For each
$\gamma\in\Gamma^{F}$ and $(X_{i},t_i)\in\mathcal{A}_{i}\times T_{i}$
with $\gamma_{i}(X_{i}|t_{i})>0$, it holds that 
\begin{equation}
X_{i}\in
\arg\max_{X_{i}'\in\mathcal{A}_{i}}\sum_{a,\,X_{-i},\,t_{-i},\,\theta}\gamma(a,X|t,\theta)\pi(t,\theta)F((X_{i}',X_{-i}),\theta),\label{GPM lemma eq 1}
\end{equation}
which, together with Definition \ref{new def: GP}, implies that 
\begin{equation}
X_{i}\cap\arg\max_{a_{i}'\in A_{i}}\sum_{a,\,X_{-i},\,t_{-i},\,\theta}\gamma(a,X|t,\theta)\pi(t,\theta)u_{i}((a_{i}',a_{-i}),\theta)\neq\emptyset.\label{GPM lemma eq 2}
\end{equation}

Fix $\hat\gamma\in\Gamma^{F}$ and let $\hat\rho:T\times\Theta\to\Delta(\mathcal{A})$
be the communication rule with $\hat\rho(X|t,\theta)=\hat\gamma(X|t,\theta)$.
Consider the conjunction of $(T,\Theta,\pi,u)$ and $\hat\rho$ and let
$\Sigma_{i}^{\mathcal{A}}$ be the set of player $i$'s strategies
in the conjunction that always assign some $a_{i}\in X_{i}$ whenever
player $i$ receives $X_{i}\in\mathcal{A}_{i}$: 
\[
\Sigma_{i}^{\mathcal{A}}=\{\sigma_{i}: T_{i}\times\mathcal{A}_{i}\to\Delta(A_{i})\mid\sigma_{i}(a_{i}|t_{i},X_{i})=0\text{ for }a_{i}\not\in X_{i}\}.
\]
For each $\sigma\in \Sigma^\calA$, an $\calA$-decision rule $\gamma$ given by $\gamma(a,X|t,\theta)=\prod_{i\in I}\sigma_{i}(a_{i}|t_{i},X_{i})\hat\rho(X|t,\theta)$ is an element of $\Gamma^F$ since $\gamma(X|t,\theta)=\hat\rho(X|t,\theta)=\hat\gamma(X|t,\theta)$.

We show that the conjunction has a BNE in which, for each $i$, player $i$ follows a strategy in $\Sigma_{i}^{\mathcal{A}}$. 
Observe that, for any $\sigma_{-i}\in\Sigma_{-i}^{\mathcal{A}}$ and $(X_{i},t_i)\in\mathcal{A}_{i}\times T_i$ with $\hat\rho_{i}(X_{i}|t_{i})>0$, it holds
that 
\[
X_{i}\cap\arg\max_{a_{i}'\in A_{i}}\sum_{a_{-i},\,X_{-i},\,t_{-i},\,\theta}\prod_{j\neq i}\sigma_{j}(a_{j}|t_{j},X_{j})\hat\rho(X|t,\theta)\pi(t,\theta)u_{i}((a_{i}',a_{-i}),\theta)\neq\emptyset
\]
by \eqref{GPM lemma eq 2}. That is, 
for any $\sigma_{-i}\in\Sigma_{-i}^{\mathcal{A}}$,
there is $\sigma_i'\in\Sigma_i^{\mathcal{A}}$ that is a best response
to $\sigma_{-i}$. 
Thus, 
the restricted best response correspondence $B^\calA:\Sigma^{\mathcal{A}}\rightrightarrows\Sigma^{\mathcal{A}}$
given by 
\begin{align*}
B^\calA(\sigma)=\{\sigma'\in\Sigma^{\mathcal{A}}\mid\text{$\sigma_i'$ is a best response strategy to $\sigma_{-i}$ for each $i\in I$}\}
\end{align*}
has nonempty convex values. 
Moreover, it can be readily shown that it has a closed graph in the distributional strategy space $\Sigma^\calA\circ \pi$. 
Applying the Kakutani--Fan--Glicksberg fixed point theorem, we obtain a BNE $\sigma^{*}\in\Sigma^{\mathcal{A}}$ of the conjunction. 
Then, the $\calA$-decision rule $\gamma^*\in \Gamma^F$ given by $\gamma^*(a,X|t,\theta)=\prod_{i\in I}\sigma_{i}^*(a_{i}|t_{i},X_{i})\hat\rho(X|t,\theta)$ is a desired BIBCE.~\hfill $\square$

\section{Proofs for Section \ref{Main results}} 

For a belief-invariant $\calA$-decision rule $\gamma\in\Gamma^{BI}$, we write $\gamma_{S}(a_{s},X_{S}|t,\theta)\equiv\gamma(\{a_{S}\}\times A_{-S}\times\{X_{S}\}\times\mathcal{A}_{-S}|t,\theta)$. 
We also write $\gamma_{i}(a_{i},X_{i}|t_{i})\equiv\gamma_{\{i\}}(a_{i},X_{i}|t,\theta)$ (since $\gamma$ is belief-invariant), $\gamma_{S}(a_{S}|t,\theta)=\gamma_{S}(\{a_{S}\}\times\mathcal{A}_{-S}|t,\theta)$, 
and $\gamma_{S}(X_{S}|t,\theta)=\gamma_{S}(A_{-S}\times\{X_{S}\}|t,\theta)$.

\subsection{Proof of Lemma \ref{main theorem lemma}}\label{Appendix for main results}

We adopt a generalized potential function $F$ satisfying the following condition. 
Let $\phi:\Theta\to\Theta^*$ be a mapping satisfying $\phi_i(\theta)=\theta_i$ if $\theta_i\in\Theta_i^*$ for all $i\in I$,  
where $\phi_i(\theta)$ is the $i$-th component of $\phi(\theta)\in \prod_{i\in I}\Theta_i^*$. 
We assume that $F$ satisfies 
\begin{equation}
F(X,\theta)=F(X,\phi(\theta)) \text{ for all }(X,\theta)\in \calA\times\Theta, \label{choice of G-potential} 	
\end{equation}
which is possible because 
the definition of generalized potential functions does not impose any conditions on their values when $\theta\not\in \Theta^*$. 

Consider $\{(T,\Theta,\pi^{k},u)\}_{k=1}^{\infty}$. For each $k$, let $T_{i}^{k}\equiv\{t_{i}\in T_{i}\mid\pi^{k}(\Theta_{i}^{*}\times\Theta_{-i}|t_{i})=1\}$
and $S^{k}(t)\equiv\{i\in I\mid t_{i}\in T_{i}^{k}\}$.
We call types in $T_i^k$ standard and all other types perturbed.

Let
\[
\Gamma^{k}\equiv\{\gamma\in\Gamma^{BI}\mid\gamma(a,X|t,\theta)=\gamma_{S^{k}(t)}(a_{S^{k}(t)},X_{S^{k}(t)}|t,\theta)\prod_{i\not\in S^{k}(t)}\gamma_{i}(a_{i},X_{i}|t_{i})\}
\]
denote the collection of belief-invariant $\calA$-decision rules
in which perturbed types receive independent signals and
choose independent actions. Note that $\Gamma^{k}$ is convex with
respect to the following convex combination: for $\gamma,\gamma'\in\Gamma^{k}$,
$\gamma''=\lambda\gamma+(1-\lambda)\gamma'\in\Gamma^{k}$ is given
by 
\[
\gamma''_{S^{k}(t)}(\cdot|t,\theta)=\lambda\gamma_{S^{k}(t)}(\cdot|t,\theta)+(1-\lambda)\gamma'_{S^{k}(t)}(\cdot|t,\theta),\ 
\prod_{i\not\in S^{k}(t)}\gamma''_{i}(\cdot|t_{i})=\prod_{i\not\in S^{k}(t)}\Big(\lambda\gamma_{i}(\cdot|t_{i})+(1-\lambda)\gamma'_{i}(\cdot|t_{i})\Big).
\]
For each $\gamma\in\Gamma^{k}$, let 
\[
\Gamma^{k}[{\gamma}]=\{\gamma'\in\Gamma^{k}\mid\gamma_{i}'(a_{i}|t_{i})=\gamma_{i}(a_{i}|t_{i})\text{ for all \ensuremath{a_{i}\in A_{i}}, \ensuremath{t_{i}\not\in T_{i}^{k}}, and \ensuremath{i\in I}}\}
\]
denote the subset of $\Gamma^{k}$ 
in which perturbed types follow $\gamma$, which is also convex.

In the next lemma, we construct a candidate for $\{\sigma^{k}\}_{k=1}^{\infty}$
satisfying \eqref{key lemma eq}.

\begin{lemma}\label{key key lemma} An $\varepsilon^{k}$-elaboration
$(T,\Theta,\pi^{k},u)$ has a BIBCE $\gamma^{k}\in\Gamma^{k}$
satisfying 
\begin{equation}
\gamma^{k}\in\arg\max_{\gamma\in\Gamma^{k}[{\gamma^{k}}]}\sum_{(X,t,\theta)\in\mathcal{A}\times T\times\Theta}\gamma(X|t,\theta)\pi^{k}(t,\theta)F(X,\theta).\label{GP-maximizer elaboration}
\end{equation}
Thus, there exists a BIBCE $\sigma^{k}\in\Sigma^{BI}$ given by $\sigma^{k}(a|t,\theta)=\gamma^{k}(a|t,\theta)$.
\end{lemma} 

Note that in a BIBCE $\gamma^k$, standard types follow vague recommendations $X_i\in \calA_i$ designed to maximize the expected value of the generalized potential function, whereas perturbed types make decisions independently.

\begin{proof}
We characterize $\gamma^{k}$ as a fixed point of a correspondence on $\Gamma^{k}$, defined by the composition of two correspondences. The existence of a fixed point is then established using a standard argument. The proof consists of three steps. 
\medskip{}

\noindent \textbf{Step 1:} The first correspondence $\Psi^{1}:\Gamma^{k}\rightrightarrows\Gamma^{k}$ is given by 
\begin{align*}
\Psi^{1}(\gamma)=\{\gamma'\in\Gamma^{k}\mid & \text{ for each \ensuremath{t_{i}\not\in T_{i}^{k}}, \ensuremath{\gamma_{i}'(a_{i}|t_{i})>0} implies }\\
 & a_{i}\in\arg\max_{a_{i}'\in A_{i}}\sum_{a_{-i},t_{-i},\theta}\gamma_{-i}(a_{-i}|t,\theta)\pi^{k}(t,\theta)u_{i}((a_{i}',a_{-i}),\theta),\\
 & \gamma'(a,X|t,\theta)=\gamma_{S^{k}(t)}(a_{S^{k}(t)},X_{S^{k}(t)}|t,\theta)\prod_{i\not\in S^{k}(t)}\gamma_{i}'(a_{i},X_{i}|t_{i})\text{ for all \ensuremath{(a,t,\theta)}}\}.
\end{align*}
Under $\gamma'\in\Psi^{1}(\gamma)$, perturbed types 
choose best responses to $\gamma$, whereas standard types 
follow~$\gamma$. 
It can be verified that $\Psi^1(\gamma)$ is nonempty and convex, and has a closed graph.
\medskip{}

\noindent \textbf{Step 2:} 
The second correspondence $\Psi^{2}:\Gamma^{k}\rightrightarrows\Gamma^{k}$
is given by 
\[
\Psi^{2}(\gamma)=\{\gamma'\in\Gamma^{k,F}[\gamma]\ \mid\text{$\gamma'$ is obedient for each $i\in I$ with $t_{i}\in T_{i}^{k}$}\},
\]
where 
\begin{align*}
\Gamma^{k,F}[\gamma]  \equiv\arg\max_{\gamma'\in\Gamma^{k}[{\gamma}]}\sum_{X,t,\theta}\gamma'(X|t,\theta)\pi^{k}(t,\theta)F(X,\theta) =\arg\max_{\gamma'\in\Gamma^{k}[{\gamma}]}\sum_{X,t,\theta}\gamma'\circ\pi^{k}(X,t,\theta)F(X,\theta).
\end{align*}
Note that \eqref{GP-maximizer elaboration} is written as $\gamma^{k}\in\Gamma^{k,F}[\gamma^{k}]$, and that $\Gamma^{k,F}[\gamma]$ is nonempty because $\{\gamma'\circ{\pi^{k}}\in\Delta(A\times\mathcal{A}\times T\times\Theta)\mid\gamma'\in\Gamma^{k}[\gamma]\}$
is tight and closed, and compact by Lemma \ref{prohorov}.
Under $\gamma'\in\Psi^{2}(\gamma)$, perturbed types follow $\gamma$, while standard
types simultaneously choose best responses. 

We show that $\Psi^{2}(\gamma)$ is nonempty by adapting the argument of Lemma \ref{Gmax BIBCE lemma}. 
Fix $\hat\gamma\in\Gamma^{k,F}[\gamma]$ and  
consider the conjunction of $(T,\Theta,\pi^k,u)$ with the communication rule 
$\hat\rho:T\times\Theta\to\Delta(\mathcal{A})$ given by $\hat\rho(X|t,\theta)=\hat\gamma(X|t,\theta)$.
Define $\Sigma_{i}^{k,\mathcal{A}}$ as the set of player $i$'s strategies
that always assign some $a_{i}\in X_{i}$ whenever
player $i$ receives $X_{i}\in\mathcal{A}_{i}$, with the mixed actions of perturbed types remaining the same as those in $\gamma$:  
\begin{align*}
\Sigma_{i}^{k,\mathcal{A}}=\{\sigma_{i}: T_{i}\times\mathcal{A}_{i}\to\Delta(A_{i})\mid\sigma_{i}(a_{i}|t_{i},X_{i})=0\text{ for }a_{i}\not\in X_{i}\text{ and } t_i\in T_i^k,\\
\quad \sigma_i(a_i|t_i,X_i)=\gamma_i(a_i|t_i)\text{ for }t_i\not\in T_i^k\}.
\end{align*}
For any $\sigma_{-i}\in\Sigma_{-i}^{k,\mathcal{A}}$ and $(X_{i},t_i)\in\mathcal{A}_{i}\times T_i^k$ with $\hat\rho_{i}(X_{i}|t_{i})>0$, \eqref{GPM lemma eq 2} implies that $X_i$ contains a best response for player $i$. 
Thus, as in the proof of Lemma \ref{Gmax BIBCE lemma}, it can be verified that 
the best response correspondence restricted to $\Sigma^{k,\calA}$ 
has nonempty convex values and a closed graph in the distributional strategy space. 
The fixed point theorem guarantees the existence of $\sigma^*\in \Sigma^{k,\calA}$ where all standard types simultaneously choose best responses. Then, the $\calA$-decision rule $\gamma^*\in \Gamma^F$ given by $\gamma^*(a,X|t,\theta)=\prod\sigma_{i}^*(a_{i}|t_{i},X_{i})\hat\rho(X|t,\theta)$ belongs to $\Psi^2(\gamma)$,  confirming that $\Psi^2(\gamma)$ is nonempty. 
It can also be verified
that $\Psi^{2}(\gamma)$ is convex and has a closed graph.
\medskip{}

\noindent\textbf{Step 3:} Define  
$\Psi:\Gamma^{k}\rightrightarrows\Gamma^{k}$ using $\Psi^{1}$ and $\Psi^{2}$:
\begin{align*}
\Psi(\gamma)=\{\gamma'\in\Gamma^{k}\mid & \,\gamma'(a,X|t,\theta)=\gamma_{S^{k}(t)}^{2}(a_{S^{k}(t)},X_{S^{k}(t)}|t,\theta)\prod_{i\not\in S^{k}(t)}\gamma_{i}^{1}(a_{i},X_{i}|t_{i})\text{ for all \ensuremath{(a,X,t,\theta)}},\\
 & \text{ where \ensuremath{\gamma^{1}\in\Psi^{1}(\gamma)} and \ensuremath{\ \gamma^{2}\in\Psi^{2}(\gamma)}}\}.
\end{align*}
If $\gamma$ is a fixed point of $\Psi$, then
$\gamma\in\Psi^{1}(\gamma)$ and $\gamma\in\Psi^{2}(\gamma)$. Thus,
$\gamma$ satisfies \eqref{GP-maximizer elaboration} and is obedient for standard types since $\gamma\in\Psi^{2}(\gamma)$. 
Moreover, it is also obedient for perturbed types because $\gamma\in\Psi^{1}(\gamma)$. Therefore, $\gamma$ is the desired BIBCE. 

To show the existence of a fixed point, consider the set
\[
\Gamma^{k}\circ{\pi^{k}}\equiv\{\gamma\circ{\pi^{k}}\in\Delta(A\times\mathcal{A}\times T\times\Theta)\mid\gamma\circ{\pi^{k}}(a,X,t,\theta)=\gamma(a,X|t,\theta)\pi^{k}(t,\theta),\ \gamma\in\Gamma^{k}\}.
\]
Since $\Gamma^{k}\circ{\pi^{k}}$ is a tight closed subset of
$\Delta(A\times\mathcal{A}\times T\times\Theta)$, it is compact by Lemma
\ref{prohorov}. 
Regard $\Psi$ as a correspondence on $\Gamma^k\circ\pi^k$ (by identifying $\Gamma^k$ with $\Gamma^k\circ\pi^k$), which inherits nonempty convex values and a closed graph from $\Psi^1$ and $\Psi^2$.
Thus, $\Psi$ has a fixed point by the Kakutani--Fan--Glicksberg fixed point theorem.
\end{proof}

Let $\gamma^{k}$ and $\sigma^{k}$ be the BIBCE of $(T,\Theta,\pi^k,u)$ constructed in Lemma~\ref{key key lemma}. 
We now prove Lemma \ref{main theorem lemma} using them. 
Define $\eta^{k}\in\Delta(A\times\mathcal{A}\times T\times\Theta)$ by 
\begin{equation}\label{eta:def}
\eta^{k}(a,X,t,\theta)\equiv\gamma^{k}\circ{\pi^{k}}(a,X,\tau^{-1}(t),\theta)=\sum_{t'\in\tau^{-1}(t)}\gamma^{k}(a,X|t',\theta){\pi^{k}}(t',\theta).
\end{equation} 
Note that $\eta^{k}(t,\theta)\equiv\sum_{(a,X)\in A\times\mathcal{A}}\eta^{k}(a,X,t,\theta)={\pi^{k}}(\tau^{-1}(t),\theta)$, which is zero if $t\not\in \tau(T)$.
We write $\eta^{k}(a,X|t,\theta)\equiv\eta^{k}(a,X,t,\theta)/\eta^{k}(t,\theta)$ 
for $(t,\theta)\in T\times\Theta$ with $\eta^{k}(t,\theta)>0$. It
can be readily shown that $\{\eta^{k}\}_{k=1}^{\infty}$ is tight,
so it has a convergent subsequence by Lemma~\ref{prohorov}, which
is denoted by $\{\eta^{k_{l}}\}_{l=1}^{\infty}$ with $\lim_{l\to\infty}\eta^{k_{l}}=\eta^{*}$.
Note that $\eta^{*}(t,\theta)
=\pi(t,\theta)$ for each $(t,\theta)\in T^{*}\times\Theta^{*}$.
Thus, we have $\eta^{*}(a,X|t,\theta)\pi(t,\theta)=\eta^{*}(a,X,t,\theta)$. 
We will regard $\eta^{*}(a,X|t,\theta)$ as an $\calA$-decision rule in $(T^{*},\Theta^{*},\pi^{*},u^{*})$.

To prove Lemma \ref{main theorem lemma}, we show
that $\{\sigma^{k}\}_{k=1}^{\infty}$ satisfies \eqref{key lemma eq},
which can be rewritten as 
\begin{equation}
\lim_{k\to\infty}\inf_{\sigma\in\mathcal{E}^{F}}\sum_{(a,t,\theta)\in A\times T\times\Theta}\left|\eta^{k}(a,t,\theta)-\sigma\circ\pi(a,t,\theta)\right|=0.\label{key lemma eq'}
\end{equation}
To establish \eqref{key lemma eq'}, it suffices to show that, for every
convergent subsequence $\{\eta^{k_{l}}\}_{l=1}^{\infty}$, there exists
$\sigma\in\mathcal{E}^{F}$ such that $\eta^{*}(a,t,\theta)=\sigma\circ\pi(a,t,\theta)$. This is because if \eqref{key lemma eq'} does not hold, then there exists
a convergent subsequence $\{\eta^{k_{l}}\}_{l=1}^\infty$ such that 
\begin{align*}
\lim_{l\to\infty}\inf_{\sigma\in\mathcal{E}^{F}}\sum_{(a,t,\theta)\in A\times T\times\Theta} & \left|\eta^{k_{l}}(a,t,\theta)-\sigma\circ\pi(a,t,\theta)\right|>0,
\end{align*}
which implies that no $\sigma\in\mathcal{E}^{F}$ satisfies $\eta^{*}(a,t,\theta)=\sigma\circ\pi(a,t,\theta)$.

We denote an arbitrary convergent subsequence 
by $\{\eta^{k}\}_{k=1}^{\infty}$ with $\lim_{k\to\infty}\eta^{k}=\eta^{*}$, instead of $\{\eta^{k_{l}}\}_{l=1}^{\infty}$, to simplify notation.
The existence of $\sigma\in\mathcal{E}^{F}$ satisfying $\eta^{*}(a,t,\theta)=
\eta^*(a|t,\theta)\pi(t,\theta)=\sigma\circ\pi(a,t,\theta)$ follows from the next three lemmas.

\begin{lemma}\label{lemmac}
The $\calA$-decision rule $\eta^{*}(a,X|t,\theta)$ in $(T^{*},\Theta^{*},\pi^{*},u^{*})$ is belief-invariant.
\end{lemma}

\begin{lemma}\label{lemmad}
The $\calA$-decision rule $\eta^{*}(a,X|t,\theta)$ in $(T^{*},\Theta^{*},\pi^{*},u^{*})$ is obedient.
\end{lemma}
\begin{lemma}\label{lemmae}
The $\calA$-decision rule $\eta^{*}(a,X|t,\theta)$ in $(T^{*},\Theta^{*},\pi^{*},u^{*})$ is an element of $\Gamma^{F}$.
\end{lemma}

\begin{proof}[Proof of Lemma \ref{lemmac}]
Fix $i\in I$ and $(t,\theta)\in T^{*}\times\Theta^{*}$. Consider sufficiently large $k$ satisfying $\varepsilon^{k}<\pi^{k}(\tau^{-1}(t))\leq \pi^{k}(\{\tau_{i}^{-1}(t_{i})\}\times T_{-i})$, which exists since $\lim_{k\to\infty}\pi^{k}(\tau^{-1}(t))=\pi(t)>0$.
Define 
\[
\zeta_{i}^{k}(a_{i},X_{i}|t_{i})\equiv\sum_{t_{i}'\in\tau_{i}^{-1}(t_{i})}\gamma_{i}^{k}(a_{i},X_{i}|t_{i}'){\pi^{k}}(t_{i}')/\pi^{k}(\tau_{i}^{-1}(t_{i})),
\]
where $\gamma_{i}^{k}(a_{i},X_{i}|t_{i}')\equiv\sum_{a_{-i},X_{-i}}\gamma^{k}(a,X|t',\theta)$,
which is well-defined by the belief-invariance of $\gamma^{k}$. 
To establish the belief-invariance of $\eta^{*}(\cdot|t,\theta)$,
it is enough to show that 
\begin{align}
\lim_{k\to\infty} & |\eta^{k}(a_{i},X_{i}|t,\theta)-\zeta_{i}^{k}(a_{i},X_{i}|t_{i})|\nonumber \\
 & =\lim_{k\to\infty}\Bigg|\sum_{t_{i}'\in\tau_{i}^{-1}(t_{i})}\gamma_{i}^{k}(a_{i},X_{i}|t_{i}')\underbrace{\left(\frac{{\pi^{k}}(\{t_{i}'\}\times\tau_{-i}^{-1}(t_{-i}),\theta)}{\pi^{k}(\tau^{-1}(t),\theta)}-\frac{{\pi^{k}}(t_{i}')}{\pi^{k}(\tau_{i}^{-1}(t_{i}))}\right)}_{\eqref{asymp BI}*}\Bigg|=0\label{asymp BI}
\end{align}
because this implies that 
$\eta^{*}(a_{i},X_{i}|t,\theta)=\lim_{k\to\infty}\zeta_{i}^{k}(a_{i},X_{i}|t_{i})$ does not depend on $t_{-i}$ and $\theta$.

To show \eqref{asymp BI}, we bound \eqref{asymp BI}*. 
By the definition of an $\varepsilon$-elaboration, there exists $T_{i}^{\flat,k}\subset T_{i}$
with $\pi^{k}(T_{i}^{\flat,k})\geq1-\varepsilon^{k}$ satisfying 
\eqref{def 3:1} and \eqref{def 3:2}. 
Then, for each $t_{i}'\in\tau_{i}^{-1}(t_i)\cap T_{i}^{\flat,k}$ 
(which is nonempty because $\pi^{k}(\tau_{i}^{-1}(t_{i}))>\varepsilon^{k}$),
it holds that $|\text{\eqref{asymp BI}*}|\leq\varepsilon^{k}{\pi^{k}}(t_{i}')C^k$, 
where $C^k=1/\pi^{k}(\tau^{-1}(t),\theta)+1/({\pi(t_{i})\pi^{k}(\tau^{-1}(t),\theta)})+1/({\pi(t_{i})\pi^{k}(\tau_{i}^{-1}(t_{i}))})$. This bound follows from a triangle-inequality decomposition\footnote{Add and subtract $\frac{\pi^k(t_i')\,\pi(t,\theta)}{\pi(t_i)\,\pi^k(\tau^{-1}(t),\theta)}$ and $\frac{\pi^k(t_i')}{\pi(t_i)}$.} applied to \eqref{asymp BI}*, with each resulting term bounded in terms of 
\eqref{def 3:1} and \eqref{def 3:2}.
Then, using $\gamma_i^k\le 1$ and the triangle inequality to bound the contribution from $t_i'\notin T_i^{\flat,k}$ in the summation,
\begin{align*}
&|\eta^k(a_i,X_i|t,\theta) - \zeta_i^k(a_i,X_i|t_i)|\\
&\le \sum_{t_i'\in\tau_i^{-1}(t_i)\cap T_i^{\flat,k}}
  \gamma_i^k(a_i,X_i|t_i')\cdot
  \varepsilon^k\,\pi^k(t_i')\cdot C^k
  \;+\; \frac{\pi^k\!\bigl(\tau_i^{-1}(t_i)\setminus T_i^{\flat,k}\bigr)}
       {\pi^k(\tau^{-1}(t),\theta)}
  \;+\; \frac{\pi^k\!\bigl(\tau_i^{-1}(t_i)\setminus T_i^{\flat,k}\bigr)}
       {\pi^k(\tau_i^{-1}(t_i))}\\[4pt]
&\le \varepsilon^k\cdot C^k
  \;+\; \frac{\varepsilon^k}{\pi^k(\tau^{-1}(t),\theta)}
  \;+\; \frac{\varepsilon^k}{\pi^k(\tau_i^{-1}(t_i))}
  \;\to\; 0
\end{align*}
as $k\to\infty$, since $C^k$, $1/\pi^k(\tau^{-1}(t),\theta)$, and $1/\pi^k(\tau_i^{-1}(t_i))$ all remain bounded. 
\end{proof}

\begin{proof}[Proof of Lemma \ref{lemmad}]
Because $\gamma^{k}$ is obedient in $(T,\Theta,\pi^{k},u)$, it holds
that 
\begin{align}
	\sum_{a_{-i},\,X_{-i}}\qty(\sum_{t_{-i},\,\theta}\gamma^{k}(a,X|t,\theta)\pi^{k}(t,\theta))u_{i}(a,\theta)\geq\sum_{a_{-i},\,X_{-i}}\qty(\sum_{t_{-i},\,\theta}\gamma^{k}(a,X|t,\theta)\pi^{k}(t,\theta))u_{i}((a_{i}',a_{-i}),\theta) \notag
\end{align}
for all $t_{i}\in T_{i}$, $X_{i}\in\mathcal{A}_{i}$, $a_{i}\in X_{i}$,
$a_{i}'\in A_{i}$, and $i\in I$. 
Summing over all $t_i\in \tau_i^{-1}(t_i')$ for each $t_i'\in \tau_i(T_i)$ and using \eqref{eta:def},  we obtain 
\begin{align}
\sum_{a_{-i},X_{-i},t_{-i}',\theta}\eta^{k}(a,X,t',\theta)u_{i}(a,\theta)\geq\sum_{a_{-i},X_{-i},t_{-i}',\theta}\eta^{k}(a,X,t',\theta)u_{i}((a_{i}',a_{-i}),\theta). \notag
\end{align}
Taking the limit of the above as $k\to\infty$ yields
\begin{align*}
\sum_{a_{-i},X_{-i},t_{-i},\theta}\eta^{*}(a,X|t,\theta)\pi(t,\theta)u_{i}(a,\theta)\geq\sum_{a_{-i},X_{-i},t_{-i},\theta}\eta^{*}(a,X|t,\theta)\pi(t,\theta)u_{i}((a_{i}',a_{-i}),\theta)
\end{align*}
since $\eta^{*}(t,\theta)=\pi(t,\theta)$. Therefore, $\eta^{*}(a,X|t,\theta)$ is obedient.
\end{proof}

\begin{proof}[Proof of Lemma \ref{lemmae}]
We show that 
\begin{equation}
\sum_{X,t,\theta}\eta^{*}(X|t,\theta)\pi(t,\theta)F(X,\theta)\geq\sum_{X,t,\theta}\hat{\gamma}(X|t,\theta)\pi(t,\theta)F(X,\theta)\label{final proof final eq}
\end{equation}
for arbitrary $\hat{\gamma}\in\Gamma^{F}$. Fix $\hat{\gamma}\in\Gamma^{F}$
and let $\hat{\gamma}^{k}\in\Gamma^{k}[\gamma^{k}]$ be such that
\[
\hat{\gamma}_{S^{k}(t)}^{k}(a_{S^{k}(t)},X_{S^{k}(t)}|t,\theta)=\hat{\gamma}_{S^{k}(t)}(a_{S^{k}(t)},X_{S^{k}(t)}|\tau(t),\theta),
\]
which is well-defined because $\hat{\gamma}$ is belief-invariant.
Note that $\hat{\gamma}^{k}(a,X|t,\theta)=\hat{\gamma}(a,X|\tau(t),\theta)$
if $t\in T^{k}$. Then, $\gamma^{k}\in\Gamma^{k,F}[\gamma^{k}]$ implies
that 
\begin{equation}
\underbrace{
\sum_{X,t,\theta}
\gamma^{k}(X|t,\theta)\pi^{k}(t,\theta)F(X,\theta)
}_{L_k}
\geq
\underbrace{
\sum_{X,t,\theta}
\hat{\gamma}^{k}(X|t,\theta)\pi^{k}(t,\theta)F(X,\theta)
}_{R_k}.
\label{final key eq 1}
\end{equation}
We show that taking the limit of \eqref{final key eq 1} yields \eqref{final proof final eq}.

Consider $L_k$. By \eqref{eta:def}, as $k\to\infty$, 
\begin{align}
L_k&=\sum_{X,t',\theta}\qty(\sum_{t\in\tau^{-1}(t')}\gamma^{k}(X|t,\theta)\pi^{k}(t,\theta))F(X,\theta)\notag =\sum_{X,t',\theta}\eta^{k}(X,t',\theta)F(X,\theta)\notag\\
& \to \sum_{X,t',\theta}
\eta^{*}(X,t',\theta)F(X,\theta)
  =\sum_{X,t',\theta}
\eta^{*}(X|t',\theta)\pi(t',\theta)F(X,\theta).
\label{final key eq 2}
\end{align}

Consider $R_k$. Then, 
\begin{align*}
R_k\geq\sum_{X,t,\theta}
\hat{\gamma}(X|\tau(t),\theta)\pi^{k}(t,\theta)F(X,\theta)+\pi^{k}(T\setminus T^{k})\qty(\inf_{X,\theta}F(X,\theta)-\sup_{X,\theta}F(X,\theta))
\end{align*}
since $\hat{\gamma}^{k}(X|t,\theta)=\hat{\gamma}(X|\tau(t),\theta)$ for all $t\in T_k$. 
The first term in the right-hand side is rewritten as follows:
\begin{align*}
 \sum_{X,t,\theta}
\hat{\gamma}(X|\tau(t),\theta)\pi^{k}(t,\theta)F(X,\theta) &= \sum_{X,t,\theta}
\hat{\gamma}(X|t,\theta)\pi^{k}(\tau^{-1}(t),\theta)F(X,\theta)\\ =\sum_{X,t,\theta}
\hat{\gamma}(X|t,\theta)\eta^{k}(t,\theta)F(X,\theta).
\end{align*}
Thus, the preceding inequality implies
\begin{align}
\lim_{k\to\infty} R_k
 & \geq\lim_{k\to\infty}\sum_{X,t,\theta}
\hat{\gamma}(X|t,\theta)\eta^{k}(t,\theta)F(X,\theta)+\pi^{k}(T\setminus T^{k})\qty(\inf_{X,\theta}F(X,\theta)-\sup_{X,\theta}F(X,\theta))\nonumber \\
 & =\sum_{X,t,\theta}
\hat{\gamma}(X|t,\theta)\eta^{*}(t,\theta)F(X,\theta)
 =\sum_{X,t,\theta}
\hat{\gamma}(X|t,\theta)\pi(t,\theta)F(X,\theta).\label{final key eq 3}
\end{align}
It follows from \eqref{final key eq 1}, \eqref{final key eq 2}, and \eqref{final key eq 3} that \eqref{final proof final eq} holds.
\end{proof}

\section{Proofs for Section \ref{Robust BIBCE of supermodular games}} 

\subsection{Proof of Proposition \ref{supermodular result}}

\label{Appendix for supermodular}

A function
$f:A\to\mathbb{R}$ is supermodular if, for any $a,b\in A$,
\[
f(a\vee b)+f(a\wedge b)\geq f(a)+f(b),
\]
where $a\vee b=(\max\{a_{i},b_{i}\})_{i\in I}$ is the join of $a$
and $b$, and $a\wedge b=(\min\{a_{i},b_{i}\})_{i\in I}$ is the meet
of $a$ and $b$. 
It is known that a potential function 
$v(\cdot,\theta):A\to\mathbb{R}$ of a supermodular game is a supermodular function
for each $\theta\in\Theta$. 

A supermodular function has the following property.

\begin{lemma}\label{supermodular result lemma 2} Let $f:A\to\mathbb{R}$
be a supermodular function. For any $\mu\in\Delta(A)$, there
exists $\mu^{*}\in\Delta(A)$ satisfying the following conditions: 
{\em (i)} The marginal probability distributions of $\mu$ and $\mu^{*}$
on $A_{i}$ are the same for each $i\in I$. 
{\em (ii)} $\sum_{a}\mu^{*}(a)f(a)\geq\sum_{a}\mu(a)f(a)$. 
{\em (iii)} The support of $\mu^{*}$ is
linearly ordered with respect to the product order $\geq_I$. 
\end{lemma} 
\begin{proof}
Without loss of generality, let $A_i=\{1,\ldots,m_i\}\subset\mathbb R$ and
regard $\mu$ as a probability distribution on $\mathbb R^I$ supported on $A$.
For each $i \in I$, let $F_i$ be the cumulative distribution function of the
marginal distribution of $\mu$ on the $i$-th coordinate. Then the quantile
function $Q_i:(0,1)\to A_i$ is given by
$Q_i(u)
  \equiv
  \inf\{x\in\mathbb R:F_i(x)\ge u\}
  =
  \min\{a_i\in A_i:F_i(a_i)\ge u\}$. 
Let $\mu^*$ be the distribution of $(Q_i(U))_{i\in I}$, where
$U$ is uniformly distributed on $(0,1)$. 

By the standard property of quantile functions, $\mu^*$ and $\mu$ have the
same marginals. Since $Q_i$ is non-decreasing for each $i$, the map
$u\mapsto (Q_i(u))_{i\in I}$ is non-decreasing with respect to the product
order. Therefore, the
support of $\mu^*$ is linearly ordered.

By construction, the random vector
$(Q_i(U))_{i\in I}$ is comonotonic (e.g., see 
\citet{dhaeneetal2002}). A standard result on comonotonic random vectors states that a comonotonic
random vector maximizes the expectation of every supermodular function among
all random vectors with the same marginals (see Theorem~2.1 of 
\citet{puccettiwang2015}, for example).
For the supermodular function $f:A\to\mathbb R$, take a bounded supermodular
extension $\bar f:\mathbb R^I\to\mathbb R$, whose existence is straightforward because $A$ is a finite product of finite chains.
Then, 
$\sum_{a\in A}\mu^*(a)f(a)=\sum_{a\in A}\mu^*(a)\bar f(a)
  \ge
  \sum_{a\in A}\mu(a)\bar f(a)=\sum_{a\in A}\mu(a)f(a)$.
\end{proof}

We are ready to prove Proposition \ref{supermodular result}. 
Let $\sigma$ be a P-maximizing BIBCE. 
By Lemma~\ref{supermodular result lemma 2}, there exists $\sigma^{*}\in\Sigma^{BI}$
that 
satisfies the following conditions for each $(t,\theta)\in T\times\Theta$: 
(i)~The marginal distributions of $\sigma(\cdot|t,\theta)$ and $\sigma^*(\cdot|t,\theta)$ on $A_i$ are the same for each $i\in I$. 
(ii) $\sum_{a}\sigma^{\ast}(a|t,\theta)v(a,\theta)\geq\sum_{a}\sigma(a|t,\theta)v(a,\theta)$. 
(iii) The support of $\sigma^{\ast}(\cdot|t,\theta)$ is linearly ordered
with respect to $\geq_I$. 

For each $i\in I$, let $\overline{a_i}(t_i)$ and $\underline{a_i}(t_i)$ denote the maximum and minimum actions in the support of $\sigma^*_{i}(\cdot|t_{i})=\sigma_{i}(\cdot|t_{i})\in \Delta(A_i)$, respectively.
By the third condition, 
the maximum and minimum action profile in the support of $\sigma^{\ast}(\cdot|t,\theta)$ must be $\overline{a}(t)\equiv (\overline{a_i}(t_i))_{i\in I}$ and $\underline{a}(t)\equiv (\underline{a_i}(t_i))_{i\in I}$, respectively. 

By the second condition, it holds that $\sigma^{*}\in\calE^v$. 
This further implies that,  
for each $(t,\theta)\in T^*\times \Theta^*$, the function $v(\cdot,\theta)$ is constant over 
the support of $\sigma^{\ast}(\cdot|t,\theta)$; that is, 
$v(\overline{a}(t),\theta)=v(\underline{a}(t),\theta)=v(a,\theta)$ for any $a$ in the support. 
Otherwise, there would exist an action profile $a$ in the support that attains the maximum value of $v(\cdot,\theta)$ which is strictly greater than that of some other action profile in the support.
In that case, we could construct a belief-invariant decision rule that achieves a strictly higher expected value of the potential function than $\sigma^*$ by assigning probability one to $a$ when $(t,\theta)$ is realized, which contradicts $\sigma^*\in \calE^v$.

Since $\overline{\sigma}(\overline{a}(t)|t,\theta)=1$ and  $\underline{\sigma}(\underline{a}(t)|t,\theta)=1$ for all $(t,\theta)\in T^*\times\Theta^*$, we have  
\[
\sum_{a,t,\theta}\sigma^{*}(a|t,\theta)\pi(t,\theta)v(a,\theta)=\sum_{a,t,\theta}\underline{\sigma}(a|t,\theta)\pi(t,\theta)v(a,\theta)=\sum_{a,t,\theta}\overline{\sigma}(a|t,\theta)\pi(t,\theta)v(a,\theta),
\]
which implies that $\underline{\sigma},\overline{\sigma}\in\calE^v$.\hfill $\square$

\section{Proofs for Section \ref{discussion section}} 

\subsection{Proof of Lemma \ref{KM equivalence}}

Suppose that $(\bar{T},\bar{\Theta},\bar{\pi},\bar{u})$ satisfies the first two conditions of Definition \ref{def: 2} for $\varepsilon>0$. 
Then, $\bar{\pi}(\theta)\geq\bar{\pi}(\bar{T}^{\sharp})=1-\varepsilon$,
and thus $|\bar{\pi}(\tau^{-1}(t),\theta)-\pi(t,\theta)|=|\bar{\pi}(\theta)-1|\leq\varepsilon\leq\sqrt{\varepsilon}$,
where $\tau_{i}:\bar{T}_{i}\to\{t\}$. Thus, (\ref{def 3:1}) holds
with $\varepsilon$ replaced by $\sqrt{\varepsilon}$. Let $\bar{T}_{i}^{\flat}=\{\bar{t}_{i}\in\bar{T}_{i}\mid\bar{\pi}(\theta|\bar{t}_{i})\geq1-\sqrt{\varepsilon}\}$.
Then, for all $\bar{t}_{i}\in\bar{T}_{i}^{\flat}$, 
\[
|\bar{\pi}(\tau_{-i}^{-1}(t_{-i}),\theta|\bar{t}_{i})-{\pi}(t_{-i},\theta|\tau_{i}(\bar{t}_{i}))|=|\bar{\pi}(\theta|\bar{t}_{i})-{\pi}(\theta|\tau_{i}(\bar{t}_{i}))|=|\bar{\pi}(\theta|\bar{t}_{i})-1|\leq\sqrt{\varepsilon},
\]
and thus (\ref{def 3:2}) holds with $\varepsilon$ replaced by $\sqrt{\varepsilon}$.
Moreover, we must have $\bar{\pi}(\bar{T}_{i}^{\flat})\geq1-\sqrt{\varepsilon}$
because 
\begin{align*}
1-\varepsilon\leq\bar{\pi}(\theta)&=\sum_{\bar{t}_{i}\in\bar{T}_{i}^{\flat}}\bar{\pi}(\theta|\bar{t}_{i})\bar{\pi}(\bar{t}_{i})+\sum_{\bar{t}_{i}\not\in\bar{T}_{i}^{\flat}}\bar{\pi}(\theta|\bar{t}_{i})\bar{\pi}(\bar{t}_{i})\\
&\leq\bar{\pi}(\bar{T}_{i}^{\flat})+(1-\sqrt{\varepsilon})(1-\bar{\pi}(\bar{T}_{i}^{\flat}))\\
&=\sqrt{\varepsilon}\bar{\pi}(\bar{T}_{i}^{\flat})+1-\sqrt{\varepsilon}.
\end{align*}
Therefore, $(\bar{T},\bar{\Theta},\bar{\pi},\bar{u})$ is a $\sqrt{\varepsilon}$-elaboration
in the sense of Definition \ref{def: 2}.~\hfill $\square$

\subsection{Proof of Lemma \ref{critical lemma}}

To introduce notation for belief opera{critical lemma}tors, we write $\mathcal{I}=2^{I}$, $\mathcal{I}_{-i}=2^{I\setminus\{i\}}$,
$\mathcal{T}_{i}=2^{T_{i}}$, $\mathcal{T}=\{E=\prod_{i\in I}E_{i}\mid E_{i}\in\mathcal{T}_{i}\}\subset2^{T}$,
and $\mathcal{T}_{-i}=\{E_{-i}=\prod_{j\neq i}E_{j}\mid E_{j}\in\mathcal{T}_{j}\}\subset2^{T_{-i}}$.
Each $E=\prod_{i\in I}E_{i}\in\mathcal{T}$ is associated with the
strategy profile $\sigma=(\sigma_{i})_{i\in I}$ given by $\sigma_{i}(1|t_{i})=1$
if $t_{i}\in E_{i}$ and $\sigma_{i}(0|t_{i})=1$ if $t_{i}\not\in E_{i}$.
For each $i\in I$, we define the ``payoff increment'' function
$f_{i}:\mathcal{I}_{-i}\times\Theta\to\mathbb{R}$ by 
\[
f_{i}(S,\theta)=u_{i}(\mathbf{1}_{S\cup\{i\}},\theta)-u_{i}(\mathbf{1}_{S},\theta)
\]
for $(S,\theta)\in\mathcal{I}_{-i}\times\Theta$, which is the payoff increment for player $i$ by switching his action from
$0$ to $1$ when the set of the opponents playing action $1$ is
$S$ and the state is $\theta$. By the supermodularity, $f_{i}(S,\theta)\leq f_{i}(S',\theta)$
whenever $S\subseteq S'$.

For $i\in I$ and $E_{-i}=\prod_{j\neq i}E_{j}\in\mathcal{T}_{-i}$,
define the function $S_{E_{-i}}:T_{-i}\to\mathcal{I}_{-i}$ by 
\[
S_{E_{-i}}(t_{-i})=\{j\mid t_{j}\in E_{j},\ j\neq i\},
\]
which is the set of the opponents whose types are in $E_{-i}$. 
The conditional expected value of $f_{i}(S_{E_{-i}}(t_{-i}),\theta)$
given $t_{i}$, 
\[
\mathrm{E}[f_{i}(S_{E_{-i}}(t_{-i}),\theta)|t_{i}]=\sum_{t_{-i},\theta}\pi(t_{-i},\theta|t_{i})f_{i}(S_{E_{-i}}(t_{-i}),\theta),
\]
is the expected payoff increment for player $i$ of type $t_{i}$
when each $j\neq i$ chooses action $1$ if $t_{j}\in E_{j}$ and
action $0$ if $t_{j}\not\in E_{j}$.

\begin{definition} For $i\in I$, $t_{i}\in T_{i}$, and $E=\prod_{j}E_{j}\in\mathcal{T}$,
type $t_{i}$ is said to have $f_{i}$-belief about $E$ if $t_{i}\in E_{i}$
and $\mathrm{E}[f_{i}(S_{E_{-i}}(t_{-i}),\theta)|t_{i}]\geq0$. Player
$i$'s $f_{i}$-belief operator $B_{i}^{f_{i}}:\mathcal{T}\to\mathcal{T}_{i}$
is defined by 
\[
B_{i}^{f_{i}}(E)=\{t_{i}\in E_{i}\mid\mathrm{E}[f_{i}(S_{E_{-i}}(t_{-i}),\theta)|t_{i}]\geq0\}
\]
for each $E=\prod_{j}E_{j}\in\mathcal{T}$; that is, $B_{i}^{f_{i}}(E)$
is the set of player $i$'s types that have $f_{i}$-belief about
$E$. 
\end{definition}

When $\Theta$ is a singleton, it is
the same as $f_{i}$-belief introduced by \citet{morrisshin2007}
and \citet{morrisetal2016}. 

If $F=\prod_{j}F_{j}\in\mathcal{T}$ satisfies $F_{i}\subset B_{i}^{f_{i}}(F)$
for each $i\in I$, we say that $F$ is $\mathbf{f}$-evident, where
$\mathbf{f}=(f_{i})_{i\in I}$. For an $\mathbf{f}$-evident event
$F$, let $\sigma=(\sigma_{i})_{i\in I}$ be the associated strategy
profile; that is, $\sigma_{i}(1|t_{i})=1$ if $t_{i}\in F_{i}$ and
$\sigma_{i}(0|t_{i})=1$ if $t_{i}\not\in F_{i}$ for each $i\in I$.
Then, it is clear that, for each $t_{i}\in F_{i}$ and $i\in I$,
action 1 is a best response to $\sigma_{-i}$.

A typical $\mathbf{f}$-evident event is given by a common $\mathbf{f}$-belief
operator, which is defined as follows. For a payoff increment function
profile $\mathbf{f}=(f_{i})_{i\in I}$ and $E=\prod_{i}E_{i}\in\mathcal{T}$,
let 
\begin{align*}
B_{i}^{\mathbf{f},0}(E) & =E_{i},\\
B_{i}^{\mathbf{f},n+1}(E) & =B_{i}^{f_{i}}\left(\prod_{j}B_{j}^{\mathbf{f},n}(E)\right)\text{ for }n\geq0,\\
CB_{i}^{\mathbf{f}}(E) & =\bigcap_{n=0}^{\infty}B_{i}^{\mathbf{f},n}(E).
\end{align*}
We say that $t_{i}\in T_{i}$ has common $\mathbf{f}$-belief about
$E\in\mathcal{T}$ if $t_{i}\in CB_{i}^{\mathbf{f}}(E)$. We write
$CB^{\mathbf{f}}(E)=\prod_{j}CB_{j}^{\mathbf{f}}(E)$ and call $CB^{\mathbf{f}}:\mathcal{T}\to\mathcal{T}$
a common $\mathbf{f}$-belief operator. The following result is a
straightforward generalization of the corresponding result for common
$\mathbf{p}$-belief and common $\mathbf{f}$-belief when $\Theta$
is a singleton. \begin{proposition} For each $E\in\mathcal{T}$,
$CB^{\mathbf{f}}(E)$ is the largest $\mathbf{f}$-evident event contained
in $E$. \end{proposition}

This proposition implies that $CB^u=CB^{\mathbf{f}}$. 

Let $E^{*}=\prod_{i}E_{i}^{*}\in\mathcal{T}$
be such that $E^{*}\subseteq E$ and 
\begin{equation}
\sum_{(t,\theta)\in T\times\Theta}\pi(t,\theta)v(\mathbf{1}_{S_{E^{*}}(t)},\theta)\geq\sum_{(t,\theta)\in T\times\Theta}\pi(t,\theta)v(\mathbf{1}_{S_{E'}(t)},\theta)\label{MP maximizing Nash}
\end{equation}
for all $E'=\prod_{i}E_{i}'\in\mathcal{T}$ with $E'\subseteq E$,
where $S_{E'}(t)=\{i\in I\mid t_{i}\in E'_{i}\}$. Then, for each
$t_{i}\in E_{i}^{*}$, it holds that 
\begin{align*}
E[f_{i}(S_{E_{-i}^{*}}(t_{-i}),\theta)|t_{i}] & =\sum_{t_{-i},\theta}\pi(t_{-i},\theta|t_{i})(u_{i}(\mathbf{1}_{S_{E_{-i}^{*}}(t_{-i})\cup\{i\}},\theta)-u_{i}(\mathbf{1}_{S_{E_{-i}^{*}}(t_{-i})},\theta))\\
 & \geq\frac{1}{\lambda_{i}}\sum_{t_{-i},\theta}\pi(t_{-i},\theta|t_{i})(v(\mathbf{1}_{S_{E_{-i}^{*}}(t_{-i})\cup\{i\}},\theta)-v(\mathbf{1}_{S_{E_{-i}^{*}}(t_{-i})},\theta))\geq0
\end{align*}
by \eqref{MP maximizing Nash}, which implies that $E^{*}$ is an
$\mathbf{f}$-evident event; that is, $E^{*}\subseteq CB^{\mathbf{f}}(E)$.
Observe that 
\begin{align*}
0\leq & \sum_{t,\theta}\pi(t,\theta)v(\mathbf{1}_{S_{E^{*}}(t)},\theta)-\sum_{t,\theta}\pi(t,\theta)v(\mathbf{1}_{S_{E}(t)},\theta)\\
= & \sum_{t\in E\setminus E^{*},\theta\in\Theta}\pi(t,\theta)(v(\mathbf{1}_{S_{E^{*}}(t)},\theta)-v(\mathbf{1},\theta))+\sum_{t\in T\setminus E,\theta\in\Theta}\pi(t,\theta)(v(\mathbf{1}_{S_{E^{*}}(t)},\theta)-v(\mathbf{1}_{S_{E}(t)},\theta))\\
\leq & -(\pi(E)-\pi(E^{*}))c_{1}+(1-\pi(E))c_{2},
\end{align*}
where 
\[
c_{1}=\inf_{S\neq I,\theta\in\Theta}v(\mathbf{1},\theta)-v(\mathbf{1}_{S},\theta)>0,\quad  
c_{2}=\sup_{S\subseteq S'\neq I,\theta\in\Theta}v(\mathbf{1}_{S},\theta)-v(\mathbf{1}_{S'},\theta),
\]
and thus
\[\pi(CB^{u}(E))=
\pi(CB^{\mathbf{f}}(E))\geq\pi(E^{*})\geq1-(1+c_{2}/c_{1})(1-\pi(E)), 
\]
which completes the proof.

\end{document}